\documentclass[preprint,12pt]{elsarticle}

\usepackage{amssymb}
\usepackage{amsmath}
\usepackage{amsfonts}
\usepackage{makeidx}
\usepackage{epstopdf}
\usepackage{color}
\usepackage{breqn}
\usepackage[colorlinks]{hyperref}
\usepackage[labelfont=bf]{caption}
\usepackage{subcaption}
\usepackage{float}
\usepackage{multirow}
\usepackage{multicol}
\usepackage{breqn}

\begin{document}
	
	\begin{frontmatter}
		
		
		
		\title{Physics-informed neural networks modeling for systems with moving immersed boundaries: application to an unsteady flow past a plunging foil}
		
		\author[a]{Rahul Sundar}

		\author[a]{Dipanjan Majumdar}
		\author[b]{Didier Lucor}
		\author[a]{Sunetra Sarkar}
		
		\affiliation[a]{organization={Department of aerospace engineering, Indian Institute of Technology Madras},
			addressline={}, 
			city={Chennai},
			postcode={600036}, 
			state={Tamil Nadu},
			country={India}}
		
		\affiliation[b]{organization={Laboratoire Interdisciplinaire des Sciences du Numérique LISN-CNRS},
			addressline={}, 
			city={Orsay},
			postcode={91403}, 
			state={},
			country={France}}

		\begin{abstract}
			Recently, physics informed neural networks (PINNs) have been explored extensively for solving various forward and inverse problems and facilitating querying applications  in fluid mechanics applications. However, work on PINNs for unsteady flows past moving bodies, such as flapping wings is scarce. Earlier studies mostly relied on transferring to a body attached frame of reference which is restrictive towards handling multiple moving bodies or deforming structures. Hence, in the present work, coupling the benefits of using a fixed Eulerian frame of reference and the capabilities of PINNs, an immersed boundary aware framework has been explored for developing surrogate models for unsteady flows past moving bodies. Specifically, simultaneous pressure recovery and velocity reconstruction from Immersed boundary method (IBM) simulation data has been investigated. 
			While, efficacy of velocity reconstruction has been tested against the fine resolution IBM data, as a step further, the pressure recovered was compared with that of an arbitrary Lagrange Eulerian (ALE) based solver. Under this framework, two PINN variants, (i) a moving-boundary-enabled standard Navier-Stokes based PINN (MB-PINN), and, (ii) a moving-boundary-enabled IBM based PINN (MB-IBM-PINN) have been formulated.
			Relaxation of physics constraints in both the models is identified to be a useful strategy in improving the predictions. A fluid-solid partitioning of the physics losses in MB-IBM-PINN has been  allowed, in order to investigate the effects of solid body points while training. This strategy enables MB-IBM-PINN to match with the performance of MB-PINN under certain loss weighting conditions. Interestingly, MB-PINN is found to be superior to MB-IBM-PINN when {\it a priori} knowledge of the solid body position and velocity are available. To improve the data efficiency of MB-PINN, a physics based data sampling technique has also been investigated. It is observed  that a suitable combination of physics constraint relaxation and physics based sampling can achieve a model performance comparable to the case of using all the data points, under a fixed training budget.
		\end{abstract}
		
		
		
		
	\end{frontmatter}
	
	
	\section{Introduction}\label{sec:Intro}
	
	Unsteady flows past flapping wings involving complex flow dynamics~\cite{lai1999jet,chin2016flapping,bose2019transition} are studied extensively to understand the underlying mechanisms of flight in natural flyers, and enable design of effective bio-mimetic propulsive~\cite{xie2022review} and energy harvesting devices~\cite{xiao2014review}. With the advancements in high performance computing, this has become possible through high fidelity computational fluid dynamics (CFD) simulations. In the context of simulating unsteady flows past moving bodies, broadly, two classes of methods exist to spatially discretise the computational domain, body conformal, and, body non-conformal mesh approaches. Arbitrary Lagrange Eulerian (ALE) framework~\cite{ sarrate2001arbitrary, farhat2001discrete} falls under the first category, where, the mesh deforms in conjunction with the body motion. An example of the latter is Immersed Boundary Method (IBM)~\cite{peskin2002immersed, kim2001immersed, mittal2005immersed}, in which a fixed Eulerian grid is used to discretise the domain where the solid boundary (described by a set of Lagrangian marker points), is immersed. Overall, IBM enables handling complex geometries and multiple moving bodies more easily by alleviating the need for mesh motion, re-meshing and the associated mesh quality constraints otherwise required in ALE. As a result, IBM incurs comparatively lesser computational and memory costs. Despite that, high fidelity simulations still involve significant computational burdens and are memory intensive for parametric exploration or real time query. These requirements are frequently encountered in solving ill-posed problems where boundary conditions are unknown, or, while solving inverse problems such as parameter estimation, system identification, optimisation~\cite{brunton2020machine}, as well as for hidden physics recovery~\cite{raissi2018hidden, pawar2020data}. 
	Hence, cost effective alternatives or surrogates to minimise the associated computational and memory costs for solving such class of problems in the domain of unsteady flows past moving bodies are necessary. 
	
	Instantaneous pressure field for flow around moving bodies such as natural swimmers or flyers is important due to the complex interactions between pressure associated with the acceleration of the body and the vortices in the flow~\cite{daniel1984unsteady}.  In experiments, either flow visualisation data of a passive scalar such as the concentration of contrast agents/smoke/dye is available~\cite{lai1999jet}, or, planar velocity data is available through particle image velocimetry (PIV) measurements~\cite{dabiri2014algorithm}. These quantities are much easier to measure than the pressure-field. Hence, recovering instantaneous pressure non-invasively from the flow visualisation/PIV velocity data would be extremely useful in understanding the underlying mechanisms and also for estimating the loads on a moving test specimen~\cite{liu2013vortex, dabiri2014algorithm, pirnia2020estimating}. In the case of numerical simulations too, near-field pressure recovery from velocity data is desirable. In some variants of IBM~\cite{lee2011sources, majumdar2020capturing}, a mathematical artifice is used to enable the computation of the aero-/hydro-dynamic loads from the total time derivative of velocity and the momentum forcing terms in the momentum conservation equation. In such cases, the pressure data are not even stored. Therefore, an accurate physics consistent near-field pressure recovery and simultaneous velocity reconstruction from the coarse/sparse velocity data would be highly desirable. This also adds to the effort towards further reducing the high memory costs, often incurred in storing high-fidelity simulation data. Indeed, it would be convenient to query both velocity and pressure once recovered at a specific spatio-temporal coordinate within the region of interest, with minimal efforts.
	
	Nevertheless, surrogate modeling based on simulation data obtained from IBM poses several challenges~\cite{balajewicz2014reduction}. Specifically, as a result of employing a fixed Eulerian background grid, the region of the grid bounded by the Lagrangian markers contains fictitious flow-field values which are necessary to the IBM computation. Thus, at any time instant, the Eulerian grid points falling within the range of the solid body motion could either lie in the fictitious flow domain (that shall be called the {\it solid} domain from now on) or the fluid domain.
	Moreover, enforcing the no-slip boundary condition on the solid boundary becomes difficult as the Lagrangian markers are disjoint from the underlying Eulerian grid points.
	Thus, while handling moving bodies or complex geometries becomes easier with IBM~\cite{kim2019immersed}, obtaining a surrogate capable of both recovering pressure and reconstructing the velocity accurately becomes challenging. 
	
	Recently, machine learning (ML) methods have gained particular interest in the fluid mechanics community towards developing surrogate models as ML is effective in solving complex inverse and ill posed problems while also suitable for real time query applications~\cite{brunton2020machine,vinuesa2022enhancing, vinuesa2022emerging,Cheng_IEEE_2023}. There are two paradigms in ML based approaches, {(i)} data driven \cite{vinuesa2022enhancing}, and, {(ii)} physics informed~\cite{brunton2020machine, willard2020integrating, karniadakis2021physics, cuomo2022scientific}. 
	Data driven approaches are well developed but require large amount of data to capture the essential dynamics of complex physical systems.
	For some cases, the combination of reduced-order modeling and machine learning allows to distill and exploit the underlying patterns and correlations within large-scale data sets, leading to enhanced predictive capabilities and streamlined simulations, ultimately accelerating the design and optimization processes~\cite{el_garroussi_tackling_2022,nony_reduced-order_2023}.
	
	But while data driven methods are physics agnostic, other methods embed partial/complete information of the underlying physics (in some manner). The present work focuses on the latter, specifically, the physics informed neural networks (PINNs)~\cite{raissi2019physics} which is a class of deep neural networks~\cite{goodfellow2016deep}. Formulated as an optimisation problem, PINNs embed prior knowledge of the governing physics through the objective/loss function~\cite{raissi2019physics}. 
	The added physics loss component regularises learning on the available data and also enables solving different forward/inverse problems with minimal change in the architecture~\cite{raissi2019physics}. In the context of inverse/ill-posed problems, PINNs are commonly used for, data-driven discovery of PDEs~\cite{raissi2019physics}, data assimilation~\cite{raissi2019deep}, and, hidden physics discovery \cite{raissi2020hidden}. Recent surveys have presented the diverse set of domains in which PINNs have been applied~\cite{karniadakis2021physics, cuomo2022scientific}. 
	Once trained on a given spatio-temporal domain, the model allows for query at any temporal/ spatial location within the given domain. Moreover, PINNs are memory efficient; once trained, the model parameters alone are sufficient to be saved to carry out the querying. Other advantages are, they are {\it automatically} differentiable by obtaining a continuous function of input variables, and extremely flexible to implement using the well developed deep learning~\cite{goodfellow2016deep} software platforms capable of automatic differentiation~\cite{baydin2017automatic} such as Tensorflow~\cite{abadi2016tensorflow}, PyTorch~\cite{paszke2019pytorch} and Jax~\cite{jax2018github}.

	Standard collocation PINNs~\cite{raissi2019physics} are sometimes difficult to train for forward problems due to underlying failure modes, broadly classified under: competing optimization objectives, and, propagation failures as discussed in ~\cite{wang2020understanding, krishnapriyan2021characterizing}.
	Various strategies have been proposed lately-- such as, loss component balancing~\cite{wang2020understanding, jin2021nsfnets}, modified backbone architectures~\cite{wang2020understanding}, adaptive sampling~\cite{wu2022comprehensive}, domain decomposition~\cite{jagtap2020conservative} and sequential learning~\cite{krishnapriyan2021characterizing, penwarden2023unified} -- to tackle these problems.
	A more detailed list of scalable algorithms to train PINNs efficiently can be found in ~\cite{shukla2022scalable, penwarden2023unified}. 
	When no solution data is available, it has been shown that PINNs are relatively easier to train~\cite{gopakumar2022loss} if sparse/coarse simulation results or partially observed experimental data are at one's disposal. It has also been shown that when simulation data is available, the expressivity of PINNs can be improved by appropriately relaxing the physics loss components, without the need for any architectural modifications~\cite{lucor2022simple}. In light of this, with simulation data being available in the present study, it is also of particular interest to investigate strategies that improve the predictive capability of PINNs under a fixed computational budget without the need for architectural changes. 
	
	PINNs based approaches have also been recently tested for problems involving moving boundaries/interfaces. A PINN was proposed~\cite{wang2021deep} with a two-network architecture to solve simple benchmark problems on free/moving interfaces (known as Stefan problems), in both forward and inverse settings. One network predicts/learns the previously unknown/known moving interface position, and the other learns the complete solution of the underlying PDEs with appropriate constraints at the interface, enforced through the loss function. 
	A two-phase incompressible flow problem with moving interface was solved in both forward and inverse settings~\cite{buhendwa2021inferring}, using PINNs with a volume of fluid formulation. Hidden physics recovery was performed using the volume fraction as training variables, where the velocity and pressure fields were recovered as hidden variables. 
	Motivated by the fictitious domain method (FDM), Yang et al.~\cite{yang2021fdm} proposed a FDM-PINN approach and demonstrated its capability in solving forward problems with stationary and moving bodies involving linear elliptic/parabolic PDEs.
	A few studies in the area of vortex induced vibrations of a bluff-body have also been reported~\cite{raissi2019deep, bai2022machine, tang2022transfer, bai2022machine}. These studies considered a body attached frame of reference to model the flow around the body. Note that transferring CFD data from an Eulerian frame of reference to a body attached frame of reference is not feasible in the case of multiple moving bodies or deforming structures. In such situations, a fixed Eulerian frame of reference (as used in IBM) is beneficial and effective. 
	To combine the advantages of IBM and PINNs, Huang et al.\cite{huang2022direct} proposed a direct forcing based IB-PINNs approach and applied it on a steady flow problem past a stationary cylinder for Reynolds numbers $Re < 40$. The physics losses were minimised in the entire domain considering the solid and fluid regions together. However, it is not clear if in the IB-PINN formulation, including the solid region in the physics loss calculation is necessary when the body position and velocity are known {\it a priori}.
	More recently, in another application on moving body~\cite{calicchia2023reconstructing}, a standard Navier-Stokes (N-S) formulation based PINN considering a non body-attached (inertial) frame of reference was used. 
	PIV data was used and the near-field pressure data around the moving body was recovered effectively in this approach.
	It would thus be interesting to test the efficacy of IB-PINN~\cite{huang2022direct} in comparison with the standard collocation based PINN in an inertial frame of reference, where the solid region is discarded. Such a comparison has not been undertaken in the literature, to the best of the authors' knowledge.
	The role of the solid region grid points in an IBM setting on the overall performance of the PINN is also yet to be understood. These are some of the pertinent issues that need to be addressed in light of the above discussion, in the context of handling flow past moving bodies especially in a non body-attached frame of reference.

	The key objectives and contributions of the present study are identified in the following. The application is in the area of unsteady incompressible flows past flapping wings and to the best of the authors' knowledge, this is the first time PINNs' performance have been analysed in this context in a body non-conformal setting. Note that PINNs based modeling of the flow in a body attached frame of reference could be restrictive and is not amenable for multiple moving bodies or deforming structures.
	A body non-conformal surrogate modeling framework using PINNs is developed here for pressure recovery from the velocity data.
	A fixed Eulerian background computational grid and a Lagrangian description of the solid boundary are considered. Given the generality of the approach, it is capable of handling multiple moving bodies/deforming structures, however, a single moving body has been considered in the present case for demonstrating the framework, for the sake of simplicity. 
	Under this framework, two variants of PINN are investigated, one based on the standard N-S equations considering the fluid region alone, and the other involving the IBM based modified N-S equations, where both fluid and solid regions are considered for physics loss computation. 
	The models are trained on coarsened velocity data and tested on high resolution data generated using an in-house IBM based unsteady flow solver~\cite{majumdar2020capturing}. 
	As discussed earlier in this section, the pressure data are not usually available/stored in the IBM approach when alternative mathematical formulation for load calculation are used~\cite{majumdar2020capturing, kim2001immersed}. Hence, demonstration of the pressure recovery is an important part of the present investigation. 
	To establish confidence, the pressure recovered using the PINN models are validated against that obtained using an ALE solver. This serves as a rigorous test of the surrogate model's capability to learn the underlying physics. To improve the model performance under a fixed computational budget, the efficacy of physics residual loss relaxation is also investigated.
	A residual loss formulation allowing relative fluid-solid relaxation is explored here, in order to resolve the role of the solid region in the performance of the second variant, when the position and the velocity of solid body are known {\it a priori}. To train the PINN models in a data efficient manner while maintaining good accuracy, a physics based training data sampling strategy is explored. 
	
	The rest of the paper is organised as follows. In section~\ref{sec:IBAPINN}, the immersed boundary aware PINNs based surrogate modeling framework is discussed in the context of pressure recovery and velocity reconstruction. The pre setup considerations for the PINNs including training-testing database generation and hyperparameter tuning are discussed in section~\ref{sec:presetup}. Sections~\ref{sec:relaxation} and \ref{sec:vssampling} demonstrate the effects of physics constraint relaxation and a physics based vorticity cut off sampling strategy. Finally, the key conclusions and further works are outlined in section~\ref{sec:conclusion}.

	\section{PINNs for unsteady flows past a moving body}\label{sec:IBAPINN}
	
	Often in flow problems involving a moving body, the solid body is allowed to move either based on a prescribed kinematics or as dictated by the structural dynamical equations coupled with the flow solver. The present study deals with the former scenario. One of the main aims of the present surrogate modelling framework is to recover the near-field pressure as a hidden variable, and reconstruct velocity data at any spatio-temporal coordinate within a domain of interest while enabling real time query. The present study proposes to achieve this by training a PINN in a body non-conformal manner.  As in the IBM, where the flow is investigated in an Eulerian grid and the solid is represented in a Lagrangian fashion, a similar strategy is proposed here and hence can also be called an immersed boundary aware (IBA) approach. This prevents the need for mapping the computational domain to a body-attached frame of reference to tackle a moving boundary problem as was done in earlier studies~\cite{raissi2019deep, bai2022machine}. 
	Under the proposed framework, two formulations, namely, (i) a moving-boundary-enabled standard Navier-Stokes based PINN (MB-PINN), and, (ii) moving-boundary-enabled IBM based PINN (MB-IBM-PINN), are studied.
	
	The details of the problem setup, unsteady flow modeling, the MB-PINN and MB-IBM-PINN architectures with their respective loss formulation and training methodology are presented in the following subsections.
	
	\subsection{The flapping foil system}
	
\begin{figure}[H]
		\centering
		\includegraphics[width = 0.8\linewidth]{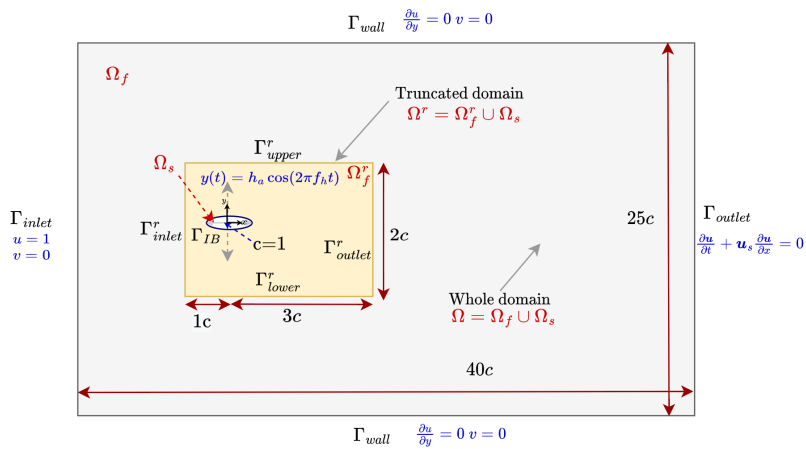}
		\caption{Schematics of the problem setup: computational domain $\Omega$ is chosen for the IBM solver, and the truncated domain $\Omega^r$ is chosen for the surrogate model. $\Omega_f$ and $\Omega_f^r$ are the fluid regions excluding the solid boundary $\Gamma_{IB}$ and solid region $\Omega_s$ at any given time instant. $\Gamma_{inlet}^r,$ $\Gamma_{outlet}^r,$ $\Gamma_{upper}^r$ and $\Gamma_{lower}^r$ are the inlet, outlet, upper and lower boundaries of the truncated domain $\Omega_r$, respectively.}
		\label{fig:problemsetup}
	\end{figure}

	Two-dimensional unsteady incompressible flow past a plunging rigid elliptic airfoil at a low Reynolds number is considered.
	The foil follows a harmonic kinematics and is considered to be of unit chord length ($c = 1$) with a thickness to chord ratio of 0.125. It is subjected to uniform free-stream; see figure~\ref{fig:problemsetup} for a schematic of the problem. 
	The plunging kinematics is given by,
	\begin{align} 
	{y}({t}) &= h_a \cos(2\pi f_h {t}), \;\; \mbox{and} \label{eq:kin1}
	\\\dot{{y}}({t}) &= -2\pi f_h h_a \sin(2\pi f_h {t}). \label{eq:kin2}
	\end{align}
	Here, $t$ is dimensional time; $h_a$ and $f_h$ are the plunging amplitude and frequency, respectively. Aligning with earlier literature~\cite{lewin2003modelling,khalid2018bifurcations}, the non-dimensional amplitude $h = h_a/c$ and reduced frequency $k = 2\pi f_h c / U_{\infty}$ are defined, where, $U_{\infty}$ is the free stream velocity. Further description of the flapping motion will be made in the terms of $k$ and $h$.
	
	\subsection{Modeling the unsteady flow}\label{sec:IBM}
	
	The flow around the flapping foil is assumed to be governed by the incompressible Navier-Stokes (N-S) equations given in its non-dimensional form by
	\begin{align}
	\frac{\partial \boldsymbol{u}}{\partial t} + \nabla.({\boldsymbol{u}}{\boldsymbol{u}}) &= -{\nabla}{p} + \frac{1}{Re} {\nabla}^2{\boldsymbol{u}}, \label{eq:NSEq1} \\
	{\nabla}.{\boldsymbol{u}}&= 0. \label{eq:NSEq2}
	\end{align}
	Here, ${\boldsymbol{u}}$ represents the non-dimensional velocity vector in the $x$-$y$ space with $u$ and $v$ being the $x$ and $y$ components of $\boldsymbol{{u}},$ respectively, and ${p}$ is the non-dimensional pressure. $\displaystyle Re = \frac{U_{\infty}c}{\nu}$ indicates the Reynolds number with $\nu$ being the kinematic viscosity.
	
\begin{figure}[t!]
		\centering
		\includegraphics[width = \linewidth]{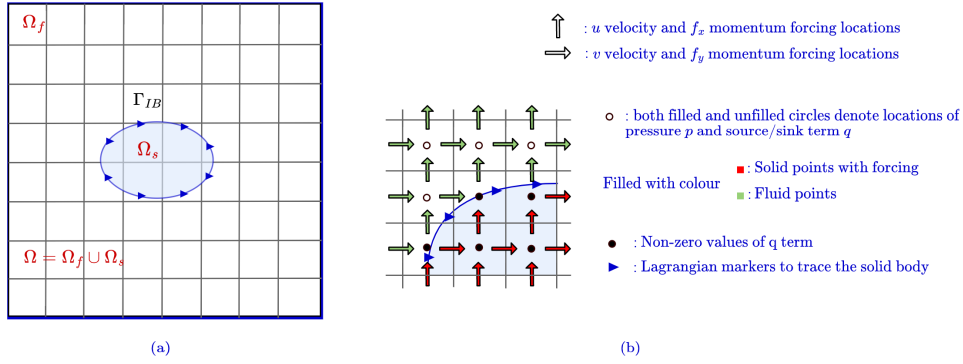}
		\caption{Schematics of (a) a representative computational domain $\Omega = \Omega_f \cup \Omega_s$ used in the immersed boundary method. Here $\Omega_s$ is the shaded area representing the solid region bounded by the solid boundary $\Gamma_{IB},$ and (b) the mesh grid representing the staggered primitive variable arrangement and fluid-solid grid point classifications. }
		\label{fig:IBM}
	\end{figure}

	A discrete forcing IBM based in-house flow solver~\cite{majumdar2020capturing} has been employed here to solve Eqs.~(\ref{eq:NSEq1}) and~(\ref{eq:NSEq2}) for generating the training and testing data for the velocity fields. In IBM approach, the immersed solid boundary $\Gamma_{IB}$ is represented by a set of Lagrangian markers (figure~\ref{fig:IBM}(a)) which do not conform exactly with the Eulerian fluid grid. Importantly, at the solid boundary $\Gamma_{IB}$, a no-slip boundary condition is to be satisfied such that $\boldsymbol{u}(\boldsymbol{x} = \Gamma_{IB},t) = \dot{y}(t)$. Since this cannot be directly enforced on $\Gamma_{IB}$ in the IBM, a momentum forcing term $\boldsymbol{f}$ with $f_x$ and $f_y$ as the respective $x$ and $y$ components must be added in Eq.~(\ref{eq:NSEq1}). $\boldsymbol{f}$ is evaluated such that the no-slip boundary condition is satisfied on $\Gamma_{IB}$ using appropriate mapping and interpolation between the Eulerian and Lagrangian points. Additionally, a source/sink term $q$ is added in Eq.~(\ref{eq:NSEq2}) to strictly satisfy the continuity equation. Notably $\boldsymbol{f}$ and $q$ are non-zero only inside $\Omega_s$ and in the vicinity of $\Gamma_{IB}$, they are zero elsewhere in $\Omega_f$ as depicted by the filled/unfilled markers and colour coded arrows in figure~\ref{fig:IBM}(b).
	The governing equations in the current IBM approach takes the form as below, 
	\begin{align}
	\frac{\partial \boldsymbol{u}}{\partial t} + \nabla.({\boldsymbol{u}}{\boldsymbol{u}}) &= -{\nabla}{p} + \frac{1}{Re} {\nabla}^2{\boldsymbol{u}} + \boldsymbol{f}, \label{eq:IBMEq1} \\
	{\nabla}.{\boldsymbol{u}} - q &= 0. \label{eq:IBMEq2}
	\end{align}
	For more details on the IBM solver, please refer to~\cite{majumdar2020capturing,shahperformance}.
	
	In the current IBM solver~\cite{majumdar2020capturing}, the pressure-field was not required for computing the aerodynamic loads and hence this data was not stored. For a rigorous testing of the pressure recovery capability of the PINN models, an ALE-based flow solver has been used to generate and test the pressure data. In the ALE technique, the governing equations are solved as it is without the need for any additional forcing or mass source/sink terms in the equations. The ALE based simulations were performed using foam-Extend's icoDyMFoam solver~\cite{jasak2007openfoam}. Details of computational domain, boundary conditions and mesh, and a thorough validation of the above IBM and ALE solvers for the low past a flapping foil can be found in the supplementary material.
	
	\subsection{Network architecture}
	
	Among the two formulations of PINNs for moving boundaries being presented here, the MB-PINN is formulated only in the fluid region (discarding the solid region) while training; the MB-IBM-PINN is formulated in both the fluid and the solid body regions. The objective is to fit velocity bulk data obtained from IBM simulations and recover pressure as a hidden variable using both MB-PINN and MB-IBM-PINN while investigating their associated efficacy.
	
\begin{figure}[!ht]
		\centering
		\begin{subfigure}{\textwidth}
			\includegraphics[width=1\linewidth]{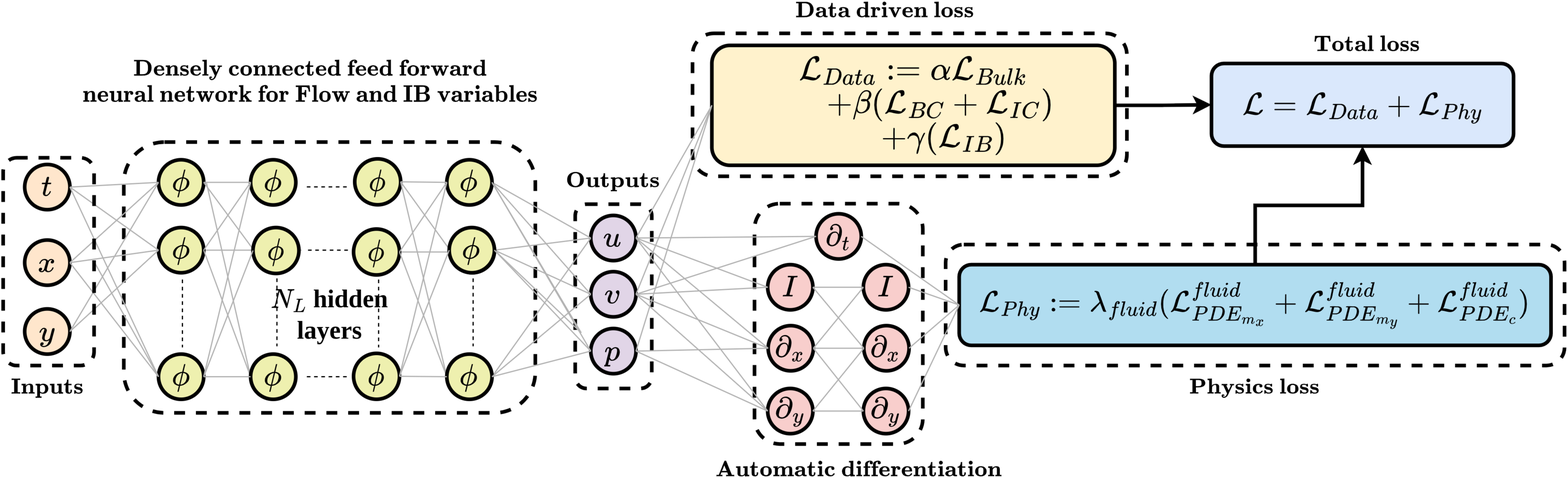}
			\caption{}
			\label{fig:ibpinndfv0}
		\end{subfigure}
		
		\begin{subfigure}{\textwidth}
			\vspace{6pt}
			\includegraphics[width=1\linewidth]{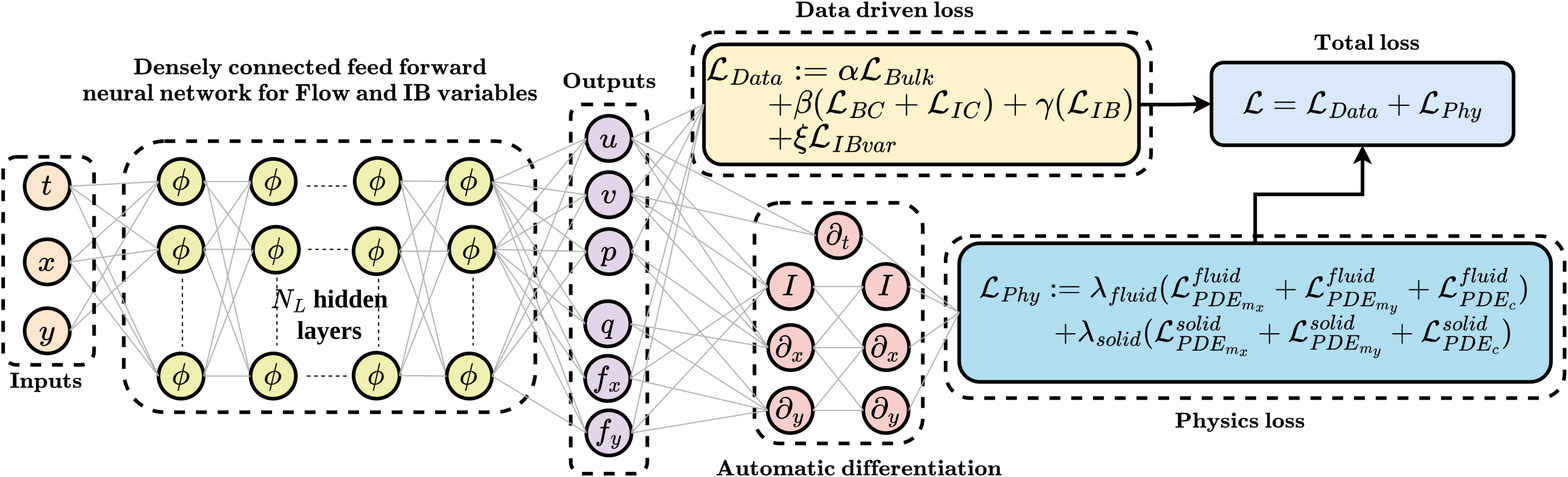}
			\caption{}
			\label{fig:ibpinndfv1}
		\end{subfigure}
		\caption{Schematic of (a) MB-PINN and (b) MB-IBM-PINN network architectures used under the present IBA framework}
		\label{fig:PINNschematic}
	\end{figure}

	For the standard collocation PINN~\cite{raissi2019physics}, a deep feed forward Neural network (FNN) is  chosen as the backbone.
	Typically, for flow problems, a FNN is formulated to approximate the flow-field variables as outputs, given a spatio-temporal input~\cite{cai2022physics}. The outputs are expressed as a composition of nonlinear activation functions acting on the spatio-temporal inputs and a network of fully connected hidden layers. Mathematically, this can be expressed as
	\begin{align}
	\boldsymbol{h}_0 &= (\boldsymbol{x},t)
	\\\boldsymbol{z}_l &= W_l\boldsymbol{h}_{l-1} + b_l,
	\\\boldsymbol{h}_l &= \phi(W_l\boldsymbol{h}_{l-1} + b_l),\;\;\text{with}\;\; l = 1, 2,\cdots L, \text{and}
	\\\boldsymbol{o} &= W_{L+1}\boldsymbol{h}_{L} + b_{L+1},
	\end{align}
	where, $(\boldsymbol{x},t)\in \mathbb{R}^{n_{\boldsymbol{x}+1}}$ is the spatio-temporal input with $\boldsymbol{h_0}\in\mathbb{R}^{n_{\boldsymbol{x}+1}}$ being the input layer and $n_{\boldsymbol{x}}$ being the spatial dimension. Here, $\boldsymbol{z}_l\in \mathbb{R}^{n_l}$ and $\boldsymbol{h}_l\in \mathbb{R}^{n_l}$ are the pre-activation and activation units of the neural network with $L$ hidden layers and $n_l$ hidden neurons for $1<l<L$, respectively, with $\phi(.)$ being a non-linear activation function. The quantities $W_l\in\mathbb{R}^{n_{l-1}\times n_l}$ and $b_l\in\mathbb{R}^{n_l}$ for $1<l<L+1$, represented by $\boldsymbol{\theta}$ as a whole, are the weights and biases of the neural network to be optimised, respectively.
	Under the IBA framework, for the MB-PINN (see schematic figure~\ref{fig:PINNschematic}(a)), the outputs $\boldsymbol{o}$ are the flow-field variables $\{\boldsymbol{u}, \; p\}$. Whereas, in MB-IBM-PINN (see schematic figure~\ref{fig:PINNschematic}(b)), the outputs $\boldsymbol{o}$ are the flow-field variables and additionally, the body forcing and mass source/sink terms such that $\boldsymbol{o} = \{\boldsymbol{u}, \; p,\; \boldsymbol{f}, \; q\}$. A FNN is often trained by optimizing the network parameters $\boldsymbol{\theta}$ with a large but potentially sparse or corrupt set of training data consisting of inputs and their corresponding target labels/outputs represented by $\{(\boldsymbol{x}^i,t^i), \boldsymbol{\hat{o}}^i\}_{i = 1}^{N_{train}}$ consisting of $N_{train}$ input-output samples with $\boldsymbol{\hat{o}}$ representing the true outputs to be fitted by the network. The optimization is carried out by minimising a loss/cost function, $\mathcal{L}$, to obtain an optimal set of network parameters, $\boldsymbol{\theta}$, such that 
	\begin{equation}
	\boldsymbol{\theta}^{optimal} = \arg \underset{\boldsymbol{\theta}}{\text{min}}{(\mathcal{L})}.
	\end{equation}
	Numerically, the loss minimisation to obtain the optimal network parameters is often performed by first randomly initialising $\boldsymbol{\theta}$ and then iteratively updating it using a variant of a stochastic gradient descent (SGD) based algorithm. Here, the gradients of $\mathcal{L}$ with respect to $\boldsymbol{\theta}$ given by $\nabla_{\boldsymbol{\theta}}\mathcal{L}$ are to be calculated. These gradients can be computed using back propagation and automatic differentiation (AD)~\cite{baydin2017automatic} which is an accurate and efficient way of computing derivatives in a computational graph following the chain rule of differentiation. AD is implemented in most modern deep learning packages, such as Tensorflow~\cite{abadi2016tensorflow} which is used in the present study.
	Generally, given $K_{MB}$ mini-batches (sub sets) of the training data containing $N_{MB}$ samples to evaluate the loss and its gradients with respect to the parameters, the algorithm for the parameter updates based on SGD is given by
	\begin{equation}
	\boldsymbol{\theta}^{i+1} \longleftarrow \boldsymbol{\theta}^{i} - \eta_i \frac{1}{N_{MB}}\sum_{i' = iN_{MB}+1}^{(i+1)N_{MB}}\nabla\mathcal{L}_{\boldsymbol{\theta}}(\boldsymbol{\theta}, (\boldsymbol{x}^{i'},t^{i'})), \;\;\;i = 1,2,\cdots K_{MB}
	\end{equation}
	where $\eta_i$ is the learning rate corresponding to $i^{th}$ epoch where each epoch corresponds to $K_{MB} = N_{train}/N_{MB}$ SGD iterations. Specifically, ADAM\cite{kingma2014adam}, a variant of SGD has been used in the literature~\cite{raissi2019physics} which will be followed in the present work as well.
	The detailed loss formulations of MB-PINN and MB-IBM-PINN are discussed in the following section.  

	\subsection{Loss formulation}
	The loss function $\mathcal{L}$ of a PINN relying on both data and physical knowledge is represented as
	\begin{align}
	\mathcal{L}:= \mathcal{L}_{Data} + \mathcal{L}_{Phy},
	\end{align}
	where $\mathcal{L}_{Data}$ and $\mathcal{L}_{Phy}$ correspond to the loss contributions from the data and physics components, respectively. Here, $\mathcal{L}_{data}$ is an indicator of the misfit between the true outputs $\hat{\boldsymbol{o}}^{train}$ and predicted network outputs $\boldsymbol{o},$ respectively; $\mathcal{L}_{Phy}$ uses AD to compute the gradients of $\boldsymbol{o}$ with respect to the input variables $(\boldsymbol{x},t)$ and compose them together to compute the governing equation residuals. This way, a PINN embeds the prior knowledge of the underlying physics in the training algorithm.
	
	The data and physics loss components can be further split into respective weighted contributions~\cite{heydari2019softadapt, wang2020understanding, bischof2021multi} from initial and boundary conditions, interior bulk data and governing equation residuals. The loss formulation in terms of the various components is as follows
	\begin{align}
	\mathcal{L}_{Data}:&= \alpha\mathcal{L}_{Bulk} + \beta (\mathcal{L}_{BC} + \mathcal{L}_{IC}) + \gamma\mathcal{L}_{IB}
	\\\mathcal{L}_{Phy}:&= \lambda(\mathcal{L}_{PDE_{m_x}} + \mathcal{L}_{PDE_{m_y}} + \mathcal{L}_{PDE_c}).
	\end{align}
	Here, $\mathcal{L}_{Bulk},\; \mathcal{L}_{BC},\; \mathcal{L}_{IC}\; \mbox{and}\; \mathcal{L}_{IB}$ are the loss components corresponding to the predicted interior bulk velocity data, velocity boundary conditions and initial condition on the Eulerian grid. The parameters $\alpha, \beta$ are the weighting coefficients of the bulk data loss, initial and boundary condition loss components, respectively. Here, $\gamma$ is the generic weighting coefficient of the additional no-slip velocity boundary condition loss $\mathcal{L}_{IB}$. It is characteristic to the IBA framework which allows enforcing the no-slip boundary conditions directly on the Lagrangian points and similar to \cite{huang2022direct} instead of following a set of sequential steps and interpolation as in IBM~\cite{majumdar2020capturing}. This is possible owing to the underlying flexibility of PINNs in choosing collocation points anywhere in the domain to enforce appropriate constraints on them. Thus, both MB-PINN and MB-IBM-PINN allow flexibility in providing both the Eulerian grid points describing the fluid and Lagrangian points describing the solid boundary, respectively. $\lambda$ is the weighting coefficient of components of the physics informed loss $\mathcal{L}_{Phy}$, which is discussed later in this section, after presenting the data driven loss components. 
	
\begin{figure}[H]
		\centering
		\includegraphics[width=0.9\linewidth]{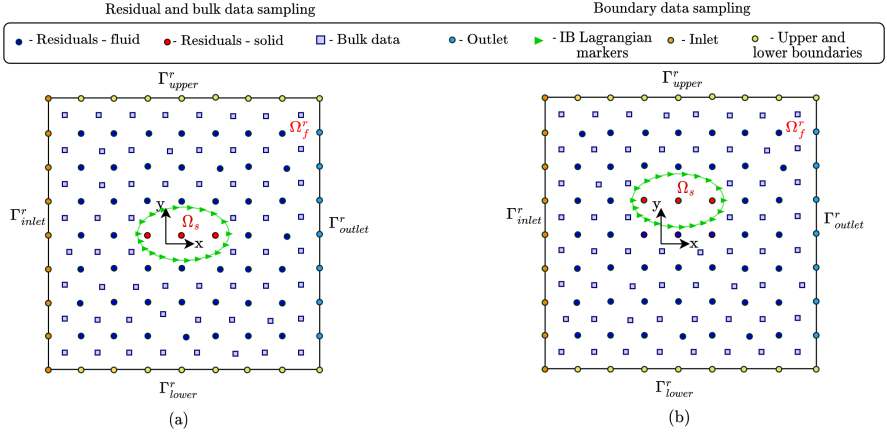}
		\caption{A schematic representing sampling of residual, bulk and boundary data points at two different time instants in (a) and (b).  Although in the present work, bulk data points and fluid region residual points are sampled from same Eulerian grid, the bulk data points and fluid residual points are shown to be disjoint considering a more general case.}
		\label{fig:schematicibmpinn}
	\end{figure}

	The data driven loss components mentioned above are common to both MB-PINN and MB-IBM-PINN, which can be mathematically expressed as
	\begin{gather}
	\mathcal{L}_{Bulk}:= \frac{1}{N_{Bulk}}\sum_{i = 1}^{N_{Bulk}}\|\boldsymbol{u}(\boldsymbol{x}_{Bulk}^i,t^i) - \hat{\boldsymbol{u}}(\boldsymbol{x}_{Bulk}^i,t^i) \|^2_{L_2},
	\\\mathcal{L}_{BC}:= \frac{1}{N_{k}}\sum_{k = 1}^{N_{k}}\left(\frac{1}{N_{BC_k}}\sum_{i = 1}^{N_{BC_k}}\|\boldsymbol{u}(\boldsymbol{x}_{BC_k}^i,t^i) - \boldsymbol{\hat{u}}(\boldsymbol{x}_{BC_k}^i,t^i) \|^2_{L_2}\right), 
	\\\mathcal{L}_{IC}:= \frac{1}{N_{IC}}\sum_{i = 1}^{N_{IC}}\|\boldsymbol{u}(\boldsymbol{x}_{IC}^i,t_0) - \boldsymbol{\hat{u}}(\boldsymbol{x}_{IC}^i,t_0) \|^2_{L_2}, \;\;\text{and}
	\\\mathcal{L}_{IB}:= \frac{1}{N_{IB}}\sum_{i = 1}^{N_{IC}}\|\boldsymbol{u}(\boldsymbol{x}_{IB}^i,t_0) - \boldsymbol{\hat{u}}(\boldsymbol{x}_{IB}^i,t_0) \|^2_{L_2},
	\end{gather}
	where, ${\boldsymbol{\hat{u}}}$ is the true velocity data generated for training by the CFD simulation, whereas, ${\boldsymbol{u}}$ is the network predicted velocity flow-field data. The spatial and temporal points are represented by $(\boldsymbol{x}, t)$ with $\boldsymbol{x} \in \Omega^r$ and $t \in [0,T]$, respectively. The subscripts shown in the loss expressions for the spatio-temporal points are to indicate independence in sampling the select points of the particular loss component (as depicted by different markers in the schematic figure~\ref{fig:schematicibmpinn}). Here, $(\boldsymbol{x}_{Bulk}^i,t^i)$ for $i = \{1,\cdots N_{Bulk}\}$ corresponds to the set of interior velocity data points excluding the solid region. The set of points $(\boldsymbol{x}_{BC_k}^i,t^i)$ for $i = \{1\cdots N_{BC_k}\}$ with $k = \{1,2,3\}$ correspond to the Dirichlet boundary conditions collected at the inlet, $\Gamma_{inlet}^r$ , upper and lower boundaries, $\Gamma_{upper}^r$ and $\Gamma_{lower}^r,$ respectively (see figure~\ref{fig:problemsetup}), of the truncated domain $\Omega^r$ considered. For simplicity, the inlet and the upper-lower boundary condition losses shall be referred to as $\mathcal{L}_{inlet}$ and $\mathcal{L}_{walls},$ respectively. The set of points $(\boldsymbol{x}_{IC}^i,t_0)$ for $i = \{1,\cdots N_{IC}\}$ are collected at the initial time stamp. Here, $(\boldsymbol{x}_{IB}^i,t^i)$ for $i = \{1,\cdots N_{IB}\}$ correspond to the set of Lagrangian markers lying on the immersed solid boundary as represented in figure~\ref{fig:schematicibmpinn}. 
	
	In the case of MB-IBM-PINN, additional constraints are required for the momentum forcing term, $\boldsymbol{f}$, and the mass source/sink term, $q$, which are also additional outputs in the network (see figure~\ref{fig:ibpinndfv1}). It is known that $\boldsymbol{f}$ and $q$ have support only in the solid region bounded by $\Gamma_{IB},$ and are zero in the fluid region at any time instant. However, both $\boldsymbol{f}$ and $q$ are not explicitly known inside the solid region as only velocity $\boldsymbol{u}$ data is stored. In~\cite{huang2022direct}, in a similar IBM based PINN formulation, velocity penalties were enforced on the Lagrangian markers to satisfy the no-slip boundary condition on the immersed boundary. It was discussed that such an enforcement compensates for the unknown $\boldsymbol{f}$ and $q$ terms in the solid region. Hence, inside the solid region, $\boldsymbol{f}$ and $q$ are recovered as hidden variables given the velocity penalties are enforced on the immersed solid boundary. This is in contrast with the discrete forcing IBM formulation~\cite{majumdar2020capturing}, where, $\boldsymbol{f}$ and $q$ are computed in a set of sequential steps to satisfy the no-slip boundary condition.
	However, since $\boldsymbol{f}$ and $q$ are known to be zero in the fluid domain, an additional data driven constraint $\mathcal{L}_{IBvar}$ is formulated based on~\cite{huang2022direct}, such that 
	\begin{gather}
	\mathcal{L}_{IBvar} := \frac{1}{N_{f}}\sum_{i = 1}^{N_{IBvar}}\left(\|f_x(\boldsymbol{x}_{IBvar}^i,t^i)\|^2_{L_2} + \|f_y(\boldsymbol{x}_{IBvar}^i,t^i)\|^2_{L_2} + \|q(\boldsymbol{x}_{IBvar}^i,t^i)\|^2_{L_2}\right),
	\end{gather}
	where, $(\boldsymbol{x}_{IBvar}^i,t^i)$ for $i = \{1,\cdots N_{IBvar}\}$ is the set of points from the interior of the fluid domain $\Omega_f^r$ where the forcing and source/sink terms loss is evaluated. 
	Thus, the data-driven loss formulation of MB-IBM-PINN is as follows  
	\begin{align}
	\mathcal{L}_{Data}:&= \alpha\mathcal{L}_{Bulk} + \beta (\mathcal{L}_{BC} + \mathcal{L}_{IC}) + \gamma\mathcal{L}_{IB} + \xi \mathcal{L}_{IBvar},
	\end{align}
	with $\xi$ being the weighting coefficient of $\mathcal{L}_{IBvar}.$\\
	
	The physics informed loss components $\mathcal{L}_{PDE_{m_x}},\; \mbox{and}\; \mathcal{L}_{PDE_{m_y}}$ correspond to the $x$ and $y$ momentum residuals and $\mathcal{L}_{PDE_c}$ corresponds to the continuity residual, respectively, which are expressed as follows
	\begin{gather}
	\mathcal{L}_{PDE_{m_x}} := \frac{1}{N_{Res}}\sum_{i = 1}^{N_{Res}}\|r_m^x(\boldsymbol{x}_{Res}^i,t^i)\|^2_{L_2},
	\\\mathcal{L}_{PDE_{m_y}} :=\frac{1}{N_{Res}}\sum_{i = 1}^{N_{Res}}\|r_m^y(\boldsymbol{x}_{Res}^i,t^i)\|^2_{L_2},\;\;\mbox{and}
	\\\mathcal{L}_{PDE_c} := \frac{1}{N_{Res}}\sum_{i = 1}^{N_{Res}}\|r_c^y(\boldsymbol{x}_{Res}^i,t^i)\|^2_{L_2}.
	\end{gather}
	For MB-PINN, the following forms of the residuals are used 
	\begin{gather}\label{eq:residuals}
	r_m^x(\boldsymbol{x}_{Res}^i,t^i) = u_t + uu_x + vu_y - p_x - 1/Re(u_{xx} + u_{yy}),	
	\\r_m^y(\boldsymbol{x}_{Res}^i,t^i) = v_t + uv_x + vv_y - p_y - 1/Re(v_{xx} + v_{yy}),\;\;\mbox{and}
	\\r_c(\boldsymbol{x}_{Res}^i,t^i)= u_x + v_y.
	\end{gather}
	In MB-IBM-PINN, the momentum and continuity residuals are expressed as
	\begin{gather}\label{eq:residualsIBM}
	r_m^x(\boldsymbol{x}_{Res}^i,t^i) = u_t + uu_x + vu_y - p_x - 1/Re(u_{xx} + u_{yy}) - f_x,	
	\\r_m^y(\boldsymbol{x}_{Res}^i,t^i) = v_t + uv_x + vv_y - p_y - 1/Re(v_{xx} + v_{yy}) - f_y,\;\;\mbox{and}
	\\r_c(\boldsymbol{x}_{Res}^i,t^i)= u_x + v_y	- q,
	\end{gather}
	where, $(\boldsymbol{x}_{Res}^i,t^i)$ for $i = \{1,\cdots N_{Res}\}$ is the set of points from the interior of the domain $\Omega^r$ where the governing equation residuals are evaluated to compute the physics loss $\mathcal{L}_{Phy}.$
	
	In MB-PINN, solid points from $\Omega_s$ at any given temporal snapshot are discarded while computing the bulk data and physics losses. This is achieved by classifying the fluid-solid points based on the solid boundary location {\it a priori} as a data preprocessing step. Thus, the set of bulk data points $(\boldsymbol{x}_{Bulk}^i, t^i)_{i = 1}^{N_{Bulk}}$ and residual collocation points $(\boldsymbol{x}_{Res}^i, t^i)_{i = 1}^{N_{Res}^{fluid}}$, only from  $\Omega_f^r$ are fed into the network while training for MB-PINN. 
	It is to be noted that unless the solid boundary points are known {\it a priori}, it is not possible to classify the domain as  solid or fluid. MB-IBM-PINN involves a more general formulation based on the governing equations (equations \ref{eq:IBMEq1} and \ref{eq:IBMEq2}), where the whole snapshot domain for the residual loss, including $\Omega_s$, is considered. For the pressure recovery problem, the boundary position and the velocity is known {\it a priori}. In such a scenario, it is of interest to understand the role of the solid region in the performance of MB-IBM-PINN and compare with MB-PINN.

	The bulk data and residual collocation points in the IBA framework come from an underlying Eulerian grid. Here, due to solid body motion, the solid region $\Omega_s$ and fluid region $\Omega_f^r$ are expected to change in time such that $\Omega^r = \Omega_f^r \cup \Omega_s.$ Thus, the underlying Eulerian grid points in the region covered by the entire range of solid body motion, can either be found in $\Omega_s$ or in $\Omega_f^r$ at any time instant as shown in \ref{fig:schematicibmpinn}(a) and \ref{fig:schematicibmpinn}(b).  Earlier works, including~\cite{huang2022direct, calicchia2023reconstructing}, do not discuss the role of solid region in the model performance. Thus, in order to determine the exact contribution from each region on the quality of predictions, a fluid-solid residual weighting is proposed in the present study for the MB-IBM-PINN formulation. Here, the residual losses are individually evaluated on the fluid and the solid regions at any time instant respectively with the corresponding loss weights. \\
	
	Thus, the $\mathcal{L}_{Phy}$ for MB-IBM-PINN is decomposed as,
	\begin{equation}\begin{split}\label{eq:part_phy1}
	\mathcal{L}_{Phy} {}=&  \lambda_{fluid}(\mathcal{L}^{fluid}_{PDE_{m_x}} + \mathcal{L}^{fluid}_{PDE_{m_y}} + \mathcal{L}^{fluid}_{PDE_c})\\ +& \lambda_{solid}(\mathcal{L}^{solid}_{PDE_{m_x}} + \mathcal{L}^{solid}_{PDE_{m_y}} + \mathcal{L}^{solid}_{PDE_c}),
	\end{split}\end{equation}
	where, $\lambda_{\#},\; \mathcal{L}^\#_{PDE_{m_x}}, \mathcal{L}^\#_{PDE_{m_y}},\text{ and }\mathcal{L}^\#_{PDE_{c}}$ correspond to the residual loss weighting coefficient, $x$ and $y$ momentum residuals and continuity residual evaluated in the respective $\#=\{\text{solid}, \text{fluid}\}$ regions. Hence, in this case, two sets of residual collocation points, one from  $\Omega_f^r$ given by $(\boldsymbol{x}_{Res}^i, t^i)_{i = 1}^{N_{Res}^{fluid}}$, and one from  region $\Omega_s$ given by $(\boldsymbol{x}_{Res}^j, t^j)_{j = 1}^{N_{Res}^{solid}}$, are needed for the evaluation of the physics losses. This allows one to sample different proportions of collocation points from each region. 
	Additionally, the relative weighting allows balancing of the relative contributions from the fluid and the solid regions. This is keeping in mind that density of the points in a particular region is equivalent to weighting the loss locally in that region~\cite{wu2022comprehensive}. 
	To make consistent comparison with the MB-IBM-PINN model and given that MB-PINN residuals are evaluated only in the fluid region, its $\mathcal{L}_{Phy}$ can be expressed in terms of $\# = \text{fluid}$ region such that
	\begin{gather}\label{eq:part_phy2}
	\mathcal{L}_{Phy}:= \lambda_{fluid}(\mathcal{L}^{fluid}_{PDE_{m_x}} + \mathcal{L}^{fluid}_{PDE_{m_y}} + \mathcal{L}^{fluid}_{PDE_c}).
	\end{gather}
	
	In every mini batch of collocation points corresponding to $\mathcal{L}_{Bulk}$ and $\mathcal{L}_{Phy}^{fluid}$ from the same fluid region $\Omega_f^r$ are mostly disjoint from each other due to random sampling. However, there still exists some overlap since the underlying grid from which these points are sampled, is kept the same in this study. Notably, given that PINNs are essentially meshless and there is freedom to select points from anywhere in the domain for the physics losses, one could also sample from a different grid than the Eulerian grid in order to keep the residual and the bulk data points  truly disjoint~\cite{wu2022comprehensive}. This allows finer control over the sampling methodology and sampling levels for each loss component individually. 
	The proportion and refinement levels for each loss component can also be varied for each individual component thereby implicitly weighting the loss components~\cite{wu2022comprehensive}. Moreover,  each individual component can be tracked and their weights can be updated adaptively, as reported in~\cite{wang2020understanding, jin2021nsfnets,wu2022comprehensive}. 
	Adaptive methods have not been investigated in the present study to focus on understanding the effect of systematic variations of loss weights and data sampling. Each loss component was monitored independently to understand how every component evolves with iterations. 
	
	The details of database generation for training and testing the models and hyperparameter tuning are discussed in section~\ref{sec:presetup}. The numerical settings considered for MB-PINN and MB-IBM-PINN along with the baseline results, and computational strategies considered to further improve the baseline performance are discussed in sections~\ref{sec:relaxation} and \ref{sec:vssampling}. 
	
	\section{Pre-setup considerations for PINN}\label{sec:presetup}
	\subsection{Training and testing database generation}\label{sec:database}
	
	Training and testing data for the velocity-field have been obtained from the IBM solver described in section~\ref{sec:IBM}. In this regard, the flow around the plunging airfoil was simulated at $Re = 500$, $k = 2\pi$ and $h = 0.16$ corresponding to $kh = 1.0$. This resulted in a periodic flow-filed with a reverse K\'arm\'an vortex street in the wake of the flapping foil. An ALE based OpenFOAM solver was employed to generate the test data set for pressure, as was mentioned in section~\ref{sec:IBM}. 
	It is known that PINNs do suffer from the issue of propagation failures~\cite{jagtap2020conservative,krishnapriyan2021characterizing}, leading to training difficulties on large spatial/temporal domains.  Moreove, earlier works~\cite{lai1999jet, platzer2008flapping, bose2018investigating, majumdar2022transition} have shown that the key flow features that largely dictate the dynamics and the aerodynamic loads in flapping problems are strongest in the vicinity of the body. 
	For the considered flapping kinematic parameters, the resulting flow is periodic~\cite{khalid2018bifurcations}, making the vortex structures and their interactions repeat in time. Keeping these in mind, to setup the training and testing data bases, a truncated spatial domain (figure~\ref{fig:problemsetup}) is chosen such that $(x,y) \in [-1c,3.5c]\times[-1c,1c]$ ensuring that at least two trailing-wake couples are contained in the truncated domain along with the flapping foil. The time domain is restricted to two plunging cycles for $t/T > 25.0$ and normalised such that $t/T\in [0,2]$ with a snapshot sampling interval of $\Delta t/T = 0.025$. Here, the snapshot interval is chosen such that the formations of the leading-edge vortex (LEV) and the trailing-edge vortex (TEV) are captured smoothly within one time period of vortex shedding. Note that the transient effects in the flow-field are over by $t/T > 25.0$ and the vortex shedding pattern is stabilised. The high resolution IBM velocity data  from a non-uniform grid is interpolated onto a coarse uniform grid to bring down the quality of the training data. This serves as a rigorous test of the IBA surrogate since the coarse grid is neither embedded in the IBM nor the ALE grid. 
	\begin{figure}[b!]
		\centering
		\includegraphics[width=0.8\linewidth]{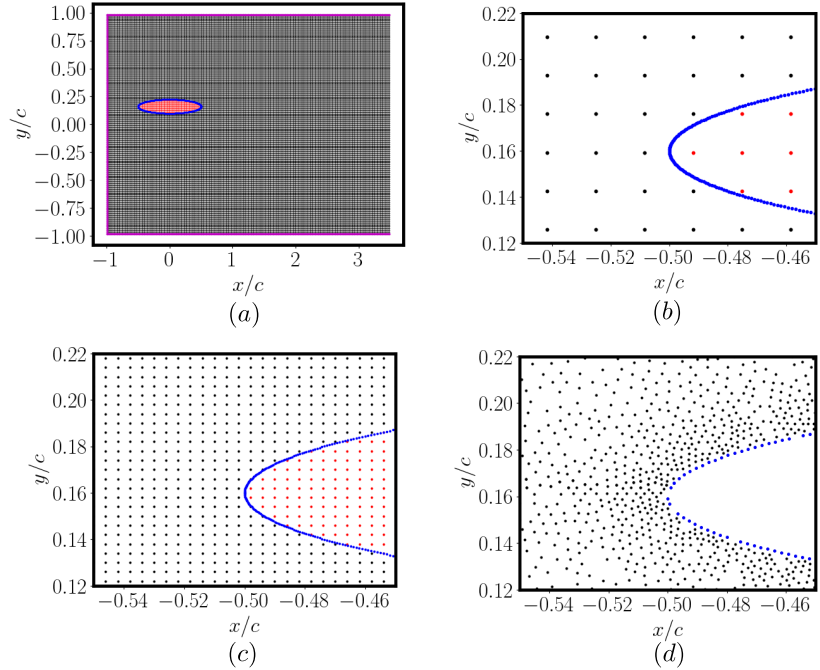}
		\caption{(a) Spatial grid of CI data set with the Lagrangian markers in blue, inlet and wall boundary points in magenta, and the solid region in red colours, respectively. Zoomed view of the spatial grid for (b) CI, (c) Ref-IBM and (d) Ref-ALE presents the comparative spatial resolution near the solid boundary.}
		\label{fig:CIdataset}
	\end{figure}
	
	Thus, three benchmark data sets are generated from the CFD simulations, (i) reference IBM velocity data (Ref-IBM), (ii) reference ALE pressure data (Ref-ALE) and (iii) coarsened and interpolated (CI) velocity data. 
	In the case of CI data set, the flow-field data from the IBM solution are interpolated onto a coarse grid containing $270\times 120$ spatial grid points with a uniform spatial resolution (figure~\ref{fig:CIdataset}(a)) approximately 5 times lower near the solid boundary than the original IBM grid. However, the resolution of CI grid is sufficient such that the key vortex structures are not lost. Secondly for training, the time interval $\Delta t/T = 0.05$ is chosen such that temporal resolution is two times coarser as compared to Ref-IBM.
	Different spatial resolutions of CI, Ref-IBM, and Ref-ALE are shown in figures~\ref{fig:CIdataset}(b)-(d), respectively. The details of the spatio-temporal resolution of these data sets are summarized in table~\ref{tab:datasets}. These data sets correspond to the truncated spatio-temporal domain and not full entire domain considered for CFD simulations. The training of the models is performed solely based on the velocity data of CI. The velocity-field reconstructions are compared with Ref-IBM, whereas the pressure recovered by the PINN models is compared with Ref-ALE. 
	
	Among the total $N_x\times N_y\times N_t = 1.3284e06$ data points in CI, $N_{Bulk} = 1.2915e06$ number of points were used to calculate $\mathcal{L}_{Bulk}$ for training after discarding the solid region. This $N_{Bulk}$ is just $\displaystyle S_{Data} = \frac{N_{Bulk}}{N_{Ref}} \times 100 = 4.89\%$ of the $N_{Ref} =N_x\times N_y\times N_t$ flow-field data points available in the Ref-IBM data set. Here, in CI data set, the boundary points resolution is such that $N_{IB} = 1000$, $N_{wall} = 540$, and $N_{inlet} = 120.$ At the initial time stamp $t/T = 0,$ $N_{IC} = 31500$ flow field data points are sampled discarding the solid region. Here, the residual collocation points are sampled from the CI grid such that there are $N_{res}^{fluid} = 1.2915e06$ and $N_{res}^{solid} = 1.476e04$ spatio-temporal points in the fluid and solid region respectively. Throughout the study, the number of boundary points and the number of residual collocation points are not varied. 
	
	By choosing the grid as in CI, there is a significant reduction in the number of training data points that encompass the flow-field information. This is in fact similar to a super-resolution setup with reconstruction of the flow-field from a coarse data~\cite{fukami2019super}. It is noted that data coarseness is often more pronounced in the super-resolution studies.  In the current study, we attempt to train the proposed PINN models only based on velocity data from the coarse CI data set to predict the velocity and pressure to be validated against Ref-IBM and Ref-ALE data sets, respectively.
	In other words, a data reconstruction problem coupled with hidden variable recovery is at hand. It is worth noting that the pressure data is not used while training throughout the study and is only recovered as a hidden variable. 
	
	\begin{table}[t!]
		\centering
		\caption{ Training and testing data set resolution within the truncated domain considered in this study. }
		\begin{tabular}{cccccc}
			\hline
			\textbf{Data sets} & $N_x$ & $N_y$ & $N_t$& $\Delta t/T$ &($Nx\times Ny\times N_t$) \\
			\hline
			Ref-IBM & 651 & 500 & 81 & 0.025 & 2.6365e07\\
			Ref-ALE & - & - & 81 & 0.025 & 4.05e06\\
			CI & 270 & 120 & 41 & 0.05 & \textbf{1.3284e06}\\
			\hline
			
		\end{tabular}
		\label{tab:datasets}
	\end{table}
	
	To evaluate the model performance, different error metrics such as root mean squared error (aRMSE), mean absolute error (aMAE), coefficient of determination (a$R^2$) and relative root mean squared error (arRMSE also called overall relative L2 error) are considered. These are computed over velocity reconstruction and pressure recovery against the reference IBM and ALE data sets (Ref-IBM and Ref-ALE), respectively. Here, the prefix ``a'' for every error metric discussed here denotes averaging of the error metrics over $u,v$ evaluated against Ref-IBM data and $p$ evaluated against Ref-ALE data. In addition, snapshot wise relative root mean squared errors (rRMSE), and the contours of normalised point-wise normalised absolute errors are presented to determine the temporal and spatial error behaviour, respectively. Here, the point-wise absolute errors are normalised with respect to the maximum absolute value of the corresponding flow-field component. Mathematical details of the above mentioned error metrics can be found in the supplementary material. Moreover, the capability of the proposed surrogate models in accurately resolving the vorticity field is investigated as well. This is crucial in the context of flow past flapping wings where the vortex structures play an important role in dictating the dynamics.

	\subsection{Hyper parameter tuning}\label{sec:hyper-param}
	To determine an appropriate number of hidden layers and hidden neurons, the underlying FNN backbone (now referred as MB-FNN) was considered. Here, MB-FNN was trained on the CI data set to reconstruct velocity fields alone in a purely data driven manner without considering the physics losses. 
	
	The number of hidden layers and hidden neurons were chosen after a grid search on combinations of $L = [6,8,10]$ layers and $n_l = [60,80,100]$ hidden neurons respectively. The resultant models were evaluated on the basis of velocity reconstruction alone on the Ref-IBM data set (accuracy details presented in table \ref{tab:hnhlaccuracy}). It was observed that with the converged $L=10$ hidden layers and $n_l = 100$ hidden neurons, remarkable relative RMSE of $0.51\%$ in $x-$component of velocity $u$ and $1.31\%$ in $y-$component of velocity $v$ were obtained (see table~\ref{tab:mb-fnn-accuracy}). This shows that MB-FNN is quite expressive. Throughout the study, these hyper-parameters are kept the same for both MB-PINN and MB-IBM-PINN models as well.
	
	For the MB-PINN model with $L = 10$ layers and $n_l = 100$ hidden neurons, $\theta = 9.1603\text{e}04$ parameters need to be calibrated while training. Whereas, due to three additional outputs in MB-IBM-PINN, $\theta = 9.1906\text{e}04$ parameters need to be calibrated while training. Unless specified, a maximum computational budget of $5e05$ iterations per training cycle in learning rate steps of $[1e-03, 5e-04, 1e-04]$ has been considered with a mini batch size of $1.5e03$, throughout the study. Training of all the models were carried out on a single NVIDIA Tesla A100 GPU card with 40GB memory and 6912 CUDA cores. Whereas, testing was carried out on a desktop grade Quadro Pascal P2200 card with 5GB memory and 1280 CUDA cores. 
	This is to investigate whether a comparable performance from MB-PINN and MB-IBM-PINN can be obtained with the same architecture and computational budget, for not only  velocity reconstruction but also pressure recovery as a hidden variable. The multi-part physics loss weighting studies have been carried out to answer this question.

	\begin{table}[H]
		\centering
		\caption{Accuracy details for the MB-FNN models (without any physical constraints) for different number of hidden layers $L$ and hidden neurons per layer $n_l$. The model is tested on the Ref-IBM data set.
		}
		\begin{tabular}{cccccc}
			\hline 
			\multirow{2}{*}{$L$} & \multirow{2}{*}{$n_l$} & \multicolumn{4}{c}{Accuracy} \\ 
			& & aRMSE & aMAE & a$R^2$ & arRMSE \\ 
			\hline
			\multirow{3}{*}{6}& 60 & 1.74e-02&	7.9e-03&	9.9929e-01&	2.47 \\ 
			& 80 & 1.455e-02&	6.3e-03&	9.9947-01&	1.26 \\ 
			& 100 &1.34e-02&	5.55e-03&	9.9956e-01&	1.16 \\ 
			\hline
			\multirow{3}{*}{8}& 60 & 1.17e-02&	5.15e-03&	9.9967e-01&	1.67 \\ 
			& 80 & 7.9e-03&	3.25e-03&	9.9984e-01&	1.13 \\ 
			& 100 & 6.75e-03&	2.55e-03&	9.9987e-01&	0.98 \\ 
			\hline
			\multirow{3}{*}{10}& 60 &9.75e-05&	4.44e-03	&9.9976e-01&	1.4\\ 
			& 80 & 7.01e-03&	2.8e-03&	9.9987e-01&	1.01 \\ 
			& \textbf{100} & \textbf{6.15e-03}&	\textbf{2.2e-03}&\textbf{	9.9989e-01}&	\textbf{0.91} \\ 
			\hline 
		\end{tabular} 
		\label{tab:hnhlaccuracy}
	\end{table}
	
	\begin{table}[H]
		\centering
		\caption{Velocity component wise accuracy of the best MB-FNN model. The model is tested on the Ref-IBM data sets.}
		\begin{tabular}{ccccc}
			\hline
			Model & \multicolumn{4}{c}{Accuracy} \\
			\hline
			MB-FNN & RMSE & MAE & $R^2$ & rRMSE \\
			$u$ & 5.8e-03 & 2.1e-03 & 9.9997e-01 &0.51 \\
			$v$ & 6.5e-03 & 2.3e-03 & 9.9981e-01 & 1.31 \\
			\hline
		\end{tabular}
		\label{tab:mb-fnn-accuracy}
	\end{table}

	\section{Role of multi-part physics loss weighting} \label{sec:relaxation}
	
	To obtain a good model performance under a fixed computational budget, an efficient strategy could be relaxing the physics loss components~\cite{buhendwa2021inferring, lucor2022simple}, or inversely, weighting the data driven loss components higher than the physics counterpart~\cite{wang2020understanding, nguyen2022fixed, jin2021nsfnets}. The former approach has been considered in the present study, where the weights of all the data driven components are taken to be unity and only the physics loss components are relaxed by lowering their appropriate weighting coefficients.  Dynamic weighting strategies have been proposed in some of the recent studies~\cite{wang2020understanding, jin2021nsfnets, shukla2022scalable, heydari2019softadapt, bischof2021multi}. In the present study, a manual tuning strategy of the residual loss weights is adopted to understand their contributions towards pressure recovery in a systematic manner. 
	
	To recall from section~\ref{sec:IBAPINN}, data sets at the solid region are discarded for computing both $\mathcal{L}_{Bulk}$ and $\mathcal{L}_{Phy}$ in MB-PINN, thus only $\lambda_{fluid}$ is in consideration. In MB-IBM-PINN, $\mathcal{L}_{Bulk}$ computation is the same as in MB-PINN, \textit{i.e.}, only the fluid region data points are considered. The solid region is  taken into account in $\mathcal{L}_{Phy}$ computation along with the fluid region, with $\lambda_{solid}$ and $\lambda_{fluid}$ as respective weights (see Eq.~(\ref{eq:part_phy1})). In the present study, a multi-part physics loss weighting strategy thus refers to the fluid-solid partitioning of the physics loss and relative weighting between them as discussed in section~\ref{sec:IBAPINN}. This choice of a fluid-solid partitioning is done to better quantify the impact of the solid region residual losses. As a result, there is an extra hyper parameter, $\lambda_{solid}$, to vary in MB-IBM-PINN. 
	
\begin{figure}[t!]
		\centering
		\includegraphics[trim = {0 2cm 0 0}, clip, width=\linewidth]{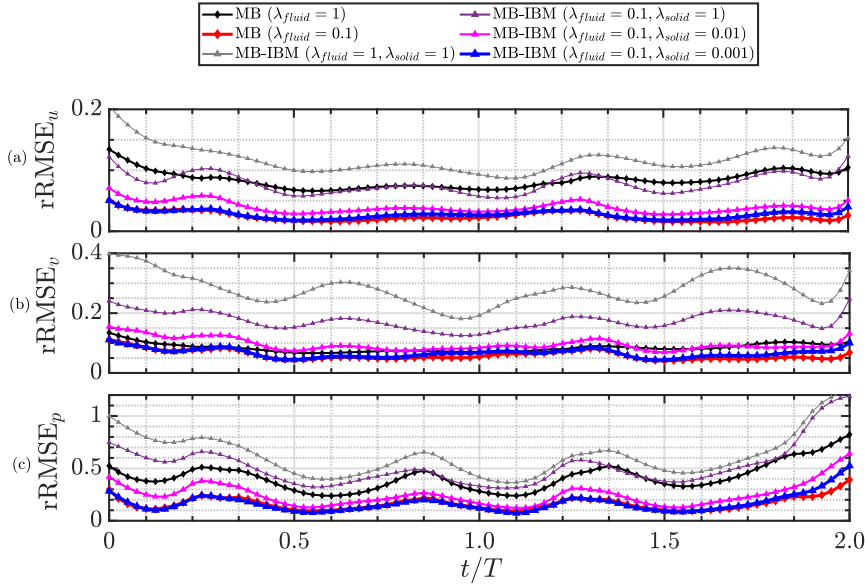}
		\caption{Relative errors (rRMSE) in time for (a, b) $x$ and $y$ velocity components with respect to Ref-IBM data set, and (c) for recovered pressure with respect to Ref-ALE data set. Here, MB-PINN (represented as MB) and MB-IBM-PINN (represented as MB-IBM) models are compared, considering $\lambda_{fluid} = 1$ and $\lambda_{fluid} = 0.1.$}
		\label{fig:relerrrelaxation}
	\end{figure}

\begin{figure}[t!]
		\centering
		\includegraphics[trim={0 13cm 0 0},clip, width=\linewidth]{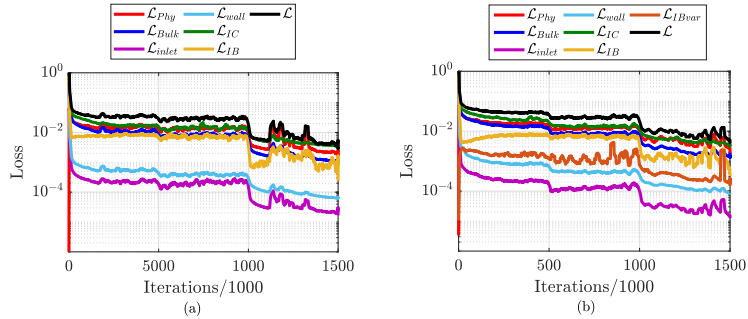}
		\caption{Smoothened individual loss component convergence shown for the baseline (a) MB-PINN with $\lambda_{fluid} = 0.1$ and (b) MB-IBM-PINN with $\lambda_{fluid} = 0.1, \lambda_{solid} = 0.001.$ }
		\label{fig:loss_conv_baseline}
	\end{figure}

	In the beginning, without any loss balancing (\textit{i.e.} $\lambda_{fluid} = \lambda_{solid} = 1.0$) it was observed that the predictions from both MB-PINN and MB-IBM-PINN  were significantly worse, with the snapshot wise relative errors being more than $20\%$ for velocity reconstruction (figure~\ref{fig:relerrrelaxation}(a)-(b)), and over $30\%$ for pressure recovery (figure~\ref{fig:relerrrelaxation}(c)). Thus, reasonable baseline models for MB-PINN and MB-IBM-PINN  were first obtained with an overall mean relative error, given by rRMSE, below $10\%$ for velocity reconstruction, and below $20\%$ for pressure recovery. This was achieved in the case of $\lambda_{fluid} = 0.1$ for MB-PINN. To match the MB-PINN's performance, for MB-IBM-PINN with the same $\lambda_{fluid} = 0.1,$ a relatively lower $\lambda_{solid} = 0.001$ was required, which suggests the need of  undermining the solid region loss. In the further discussions, the MB-PINN and MB-IBM-PINN models with $\lambda_{fluid} = 0.1$, and $\lambda_{solid} = 0.001$, shall be referred to as the baseline cases. 
	In the relative error plots (figure~\ref{fig:relerrrelaxation}), the MB-PINN and MB-IBM-PINN curves almost overlap each other except at a few later time instances, where MB-IBM-PINN does slightly worse for the selected baseline $\lambda_{fluid}$ and $\lambda_{solid}$ values. The equivalence of MB-PINN and MB-IBM-PINN is also confirmed by the similar order of the convergence of the overall training and individual loss components $\mathcal{L}_{Bulk},\; \text{scaled }\mathcal{L}_{Phy},\; \mathcal{L}_{IC},\;\mathcal{L}_{inlet},$ and $\mathcal{L}_{wall}$ in figures \ref{fig:loss_conv_baseline}(a) and \ref{fig:loss_conv_baseline}(b), respectively. 
	
\begin{figure}[t!]
		\centering
		\includegraphics[width=\linewidth]{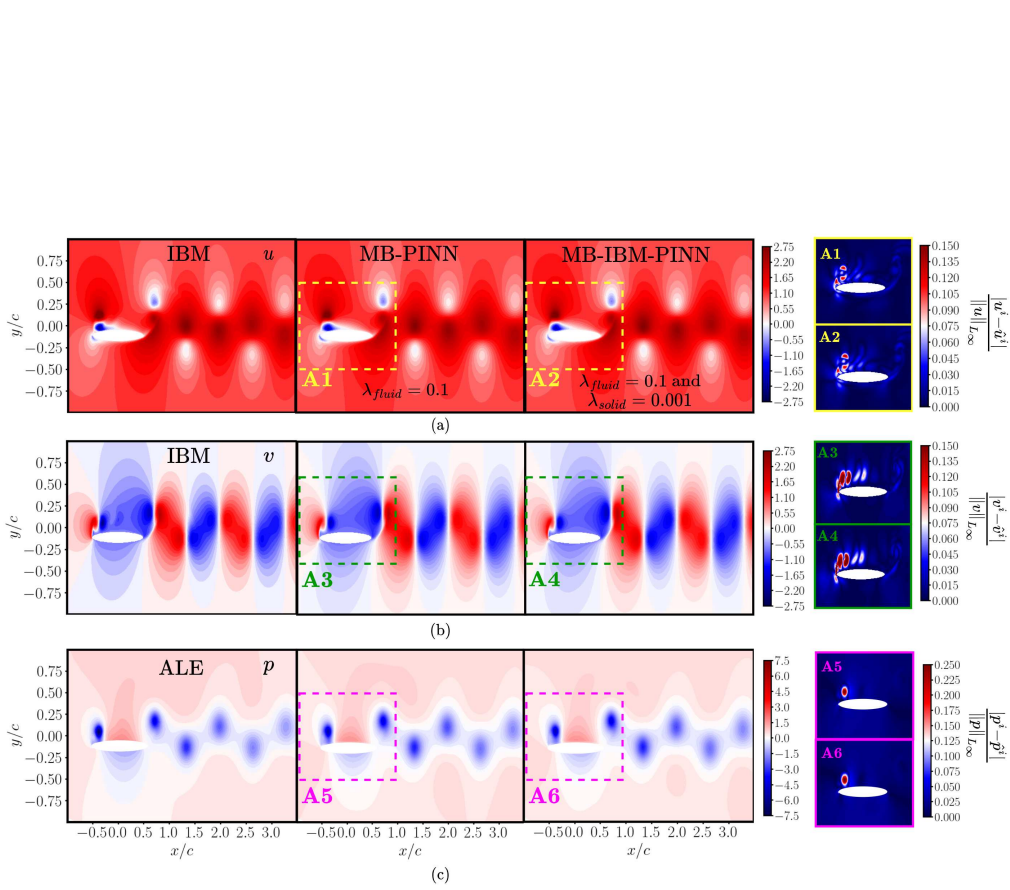}
		\caption{Comparison of true and predicted (a, b) velocity  and (c) pressure snapshots queried on the Ref-IBM and Ref-ALE grids, respectively, for a test time stamp t/T = 0.375. The inshot boxes $\textbf{A\#}$ for $\# = 1,2,\cdots 6$ correspond to the near-field region, where, the normalised point-wise absolute errors in the flow-field are prominent. }
		\label{fig:baselineUVPerrcontours}
	\end{figure}

\begin{figure}[t!]
		\centering
		\includegraphics[width=0.8\linewidth]{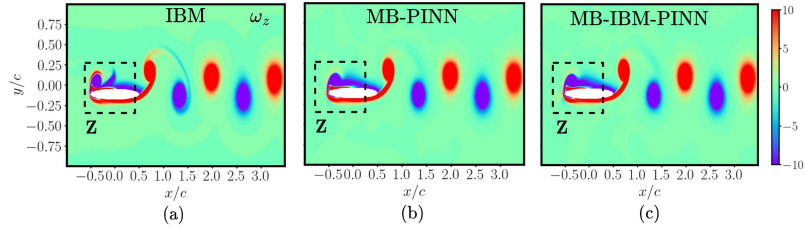}
		\caption{Comparison of (a) IBM obtained vorticity snapshot with the predictions of (b) baseline MB-PINN ($\lambda_{fluid} = 0.1$) and (c) MB-IBM-PINN ($\lambda_{fluid} = 0.1, \lambda_{solid} = 0.001$) for a test time stamp t/T = 0.375. The inshot boxes $\textbf{Z}$ correspond to the near-field region surrounding the LEV and the secondary vorticity structure.}
		\label{fig:baselinevorticitycontours}
	\end{figure}

	The true and predicted velocity and pressure contours are compared in figures~\ref{fig:baselineUVPerrcontours}(a) and \ref{fig:baselineUVPerrcontours}(b). The  corresponding maximum value normalised point-wise absolute error contours in the near-field region are shown in  figure~\ref{fig:baselineUVPerrcontours}(c) for a test time stamp $t/T = 0.375$, which is previously unseen during the training. At this time stamp, the velocity and pressure fields are queried on high resolution Ref-IBM and Ref-ALE data sets test grids, respectively. It can be seen in figure~\ref{fig:baselineUVPerrcontours}(c) that the errors are highly pronounced in the region where there is an LEV. This is expected as strong gradients of velocity and pressure exist in such regions of strong vorticity. The issue of difficulty of PINN in capturing strong/sharp gradient has also been reported in the earlier literature~\cite{wang2020understanding}. 
	For the test time $t/T = 0.375,$ 
	a characteristic strong LEV on the upper surface of the foil  followed by a secondary vortex structure in the mid-region of the upper surface is observed (see figure~\ref{fig:baselinevorticitycontours}(a)). The baseline models are able to resolve the primary vortex structures very well, while the secondary structure is still blurred in the predictions as seen in figures~\ref{fig:baselinevorticitycontours}(b) and \ref{fig:baselinevorticitycontours}(c). With regard to the pressure recovery, although the baseline results seem promising, the errors are still high. In order to see if larger training iterations ($N_{iter}$) per cycle could improve the results, in another set of experiments  with $\lambda_{fluid} = 0.1$ and $\lambda_{solid} = 0.001$ (see table~\ref{tab:longer_training_accuracy} for accuracy details), it is seen that the improvements are not too significant considering the increased computational budget. It remains to be seen if further relaxation of the residual losses can improve the pressure recovery and the overall model performance.
	
	\begin{table}[h!]
		\centering
		\caption{Accuracy details of the MB-PINN and MB-IBM-PINN models under the effect of larger $N_{iter}$.}
		\begin{tabular}{cccccc}
			\hline
			Model &$N_{iter}$ & \multicolumn{4}{c}{Accuracy} \\
			\hline
			\multirow{3}{3cm}{\centering MB-PINN ($\lambda_{fluid} = 0.1$)}& & aRMSE & aMAE & a$R^2$& arRMSE\\
			& 7.5e05 & 5.64e-02	&3.02e-02	&9.95e-01	&5.39  \\
			& 1e06& 5.24e-02	&2.79e-02	&9.96e-01	&5.01 \\
			& & & & & \\
			\multirow{3}{3cm}{\centering MB-IBM-PINN ($\lambda_{fluid} = 0.1$, $\lambda_{solid} = 0.001$)}& & &  & & \\
			& 7.5e05 & 8.79e-02&	4.87e-02	&9.87e-01	&8.32  \\
			& 1e06& 8.19e-02&	4.54e-02	&9.89e-01	&7.74  \\
			& & & & & \\
			\hline
		\end{tabular}
		\label{tab:longer_training_accuracy}
	\end{table}
	
\begin{figure}[t!]
		\centering
		\includegraphics[trim={0 2cm 0 0},clip, width=\linewidth]{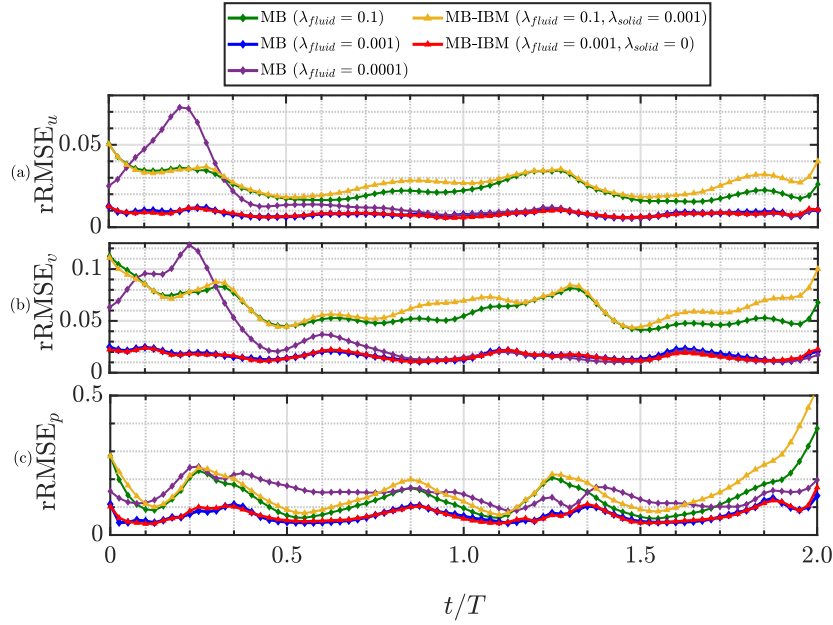}
		\caption{Relative errors (rRMSE) in time for (a,b) $x$ and $y$ velocity components, and (c) pressure $p$ predicted by MB-PINN (represented as just MB) and MB-IBM-PINN (represented as MB-IBM) models with different $\lambda_{fluid}$ and $\lambda_{solid}$ combinations depicting the effect of further residual relaxation.}
		\label{fig:relerrorbreakingpoint2}
	\end{figure}

	The detailed error measures for the relaxation studies are presented in table~\ref{tab:mb-ns-ibm-pinn-accuracy}. In the case of MB-PINN, there is a significant decrease in the overall error measures as $\lambda_{fluid}$ is decreased from $1$ to $0.1$.
	However, beyond $\lambda_{fluid} = 0.001,$   the pressure recovery breaks down as seen  in table \ref{tab:mb-ns-ibm-pinn-accuracy}. It is observed that at $\lambda_{fluid} = 0.0001$ the snapshot-wise relative errors (see figure~\ref{fig:relerrorbreakingpoint2}) are either as good as the baseline case or even worse in certain situations.   Thus the most optimal results have been obtained for $\lambda_{fluid} = 0.001$ in the case of MB-PINN, with $\text{arRMSE} = 3.22\%$ which can be considered as the best performing MB-PINN model within the present set of experiments.

	\begin{figure}[t!]
		\centering
		\includegraphics[trim={0 11cm 0 0},clip, width=\linewidth]{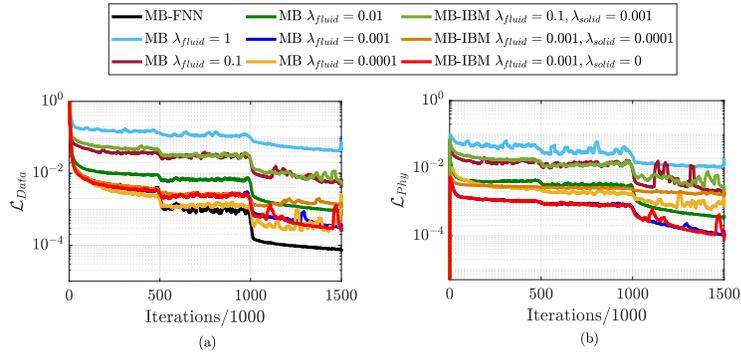}
		\caption{Overall effect of relaxation on loss convergence studied for MB-PINN and MB-IBM-PINN models, considering different $\lambda_{fluid}$ and $\lambda_{solid}$ values. The combined data driven losses $\mathcal{L}_{Data}$ are presented in comparison with the MB-FNN model in (a), and the scaled physics-informed losses $\mathcal{L}_{Phy}$ in (b), respectively.}
		\label{fig:loss_conv_relax}
	\end{figure}
	
	\begin{table}[b!]
		\centering
		\caption{Accuracy of MB-PINN and MB-IBM-PINN for different relaxation coefficients. Best performing models are highlighted in bold-faced fonts. }
		\vspace{6pt}
		\resizebox{\textwidth}{!}{\begin{tabular}{ccccccc}
				\hline
				Model & & & \multicolumn{4}{c}{Accuracy} \\
				\hline
				\multirow{5}{*}{MB-PINN} & $\lambda_{fluid}$ & $\lambda_{solid}$ & aRMSE & aMAE & a$R^2$ & arRMSE (in $\%$) \\
				\hline
				& 1 & - & 2.38e-01 & 1.37e-01 & 9.23e-01 & 22.96 \\
				& 0.1 & - &7.64e-02	&4.09e-02	&9.91e-01	&7.27 \\
				& 0.01 & - & 4.09e-02	&2.28e-02	&9.97e-01&	3.78 \\
				& \textbf{0.001} & - & \textbf{3.58e-02}	&\textbf{2.07e-02}	&\textbf{9.98e-01}	&\textbf{3.22}\\
				& 0.0001 & - & 7.59e-02&	4.87e-02&	9.91e-01	&6.56 \\
				
				\hline
				\multirow{12}{*}{MB-IBM-PINN} & $\lambda_{fluid}$ & $\lambda_{solid}$ & aRMSE & aMAE & a$R^2$ & arRMSE (in $\%$) \\
				\hline
				
				& 1 & 1 & 0.3.61e-01&	2.22e-01&	8.24e-01&	34.11 \\
				& 1 & 0.1 & 2.60e-01&	1.52e-01&	9.04e-01 &24.81 \\
				& 1 & 0.01 & 2.25e-01 &	1.32e-01	&9.28e-01&	21.56 \\
				& 1 & 0.001 & 2.08e-01	& 1.19e-01	&9.41e-01	&19.99 \\
				& 0.1 & 1 &2.83e-01&	1.71e-01&	8.85e-01&	25.76\\
				& 0.1 & 0.1 & 1.76e-01	&9.94e-02&	9.54e-01	&16.36 \\
				& 0.1 & 0.01 & 1.35e-01 	&7.60e-02&	9.73e-01&	12.66 \\
				& 0.1 & 0.001 & 9.14e-02 &	5.07e-02	& 9.87e-01	& 8.63\\
				& 0.001 & 0.001 &4.15e-01	&2.69e-01	&7.27e-01&	32.41 \\
				& 0.001 & 0.0001 & 3.27e-01&	2.078e-01 & 8.22e-01 &	25.91\\
				& 0.001 & 0.00001 & 2.71e-01&	1.719e-01 & 8.76e-01 &	21.02\\
				&\textbf{ 0.001} & \textbf{0} &\textbf{3.59e-02}&	\textbf{2.03e-02}&	\textbf{9.98e-01}	&\textbf{3.21}\\
				\hline
		\end{tabular}}
		\label{tab:mb-ns-ibm-pinn-accuracy}
	\end{table}
	For MB-IBM-PINN, even for a fixed $\lambda_{fluid},$ the prediction accuracy improves overall when the $\lambda_{solid}$ is lowered. This is observed for different $\lambda_{fluid}$ values as shown in table~\ref{tab:mb-ns-ibm-pinn-accuracy}. To obtain a reasonable accuracy in velocity reconstruction and pressure recovery, indicated by arRMSE, $\lambda_{solid}$ is expected to be at least one order lower than that of $\lambda_{fluid}.$ This is true for the baseline $\lambda_{fluid} = 0.1,$ with $\lambda_{solid} = 0.001$ that is two orders lower, where the accuracy of the model almost matches MB-PINN model for same $\lambda_{fluid}$ as shown in the relative error plots in figures~\ref{fig:relerrrelaxation}(a)-(c). This requirement could be because relaxation is also implicitly related to choosing a relatively lower proportion of collocation points in the region~\cite{wu2022comprehensive}. This is not applicable in the case of MB-PINN since the solid region is not included in the formulation. 
	By fixing $\lambda_{fluid} = 0.001$ as in the optimal MB-PINN model,  the percentage arRMSE blows up even at $\lambda_{solid} = 0.0001$ and $0.00001.$  Given that a significantly lower $\lambda_{solid}$ is indeed required to improve the accuracy further, the need for solid region losses in MB-IBM-PINN is hence investigated by choosing a limiting $\lambda_{solid}=0$  keeping $\lambda_{fluid} = 0.001.$ 
	In this case, the MB-IBM-PINN model gives the best performance, matching that of MB-PINN. The arRMSE drastically drops as seen in table~\ref{tab:mb-ns-ibm-pinn-accuracy}. 
	
	It can also be seen that for the optimal MB-PINN and MB-IBM-PINN models, $\mathcal{L}_{Data}$ (figure~\ref{fig:loss_conv_relax}(a)) comes closer to that of purely data driven MB-FNN model. Whereas, the scaled  $\mathcal{L}_{Phy}$ are the lowest for both variants of PINN at the optimal values of relaxation (see figure~\ref{fig:loss_conv_relax}(b)). 
	The relative error in time for velocity reconstruction and pressure recovery are plotted across different selected levels of $\lambda_{fluid}$ and $\lambda_{solid}$ for both PINN  models in figures \ref{fig:relerrorbreakingpoint2}(a)-(c). 
	It is seen that for MB-IBM-PINN, the relative errors in velocity reconstruction and pressure recovery closely follow the optimal MB-PINN model at all time instants for $\lambda_{fluid} =0.001$ and $\lambda_{solid} = 0$.

	It is thus seen that MB-IBM-PINN can match the performance of MB-PINN under certain combinations of $\lambda_{fluid}$ and $\lambda_{solid}$, under the same training budget and network architecture. At optimal $\lambda_{fluid}$ and for $\lambda_{solid} = 0,$ the solid region points can be effectively discarded in MB-IBM-PINN, for an equivalent performance to MB-PINN. This is because the no-slip boundary condition in both the PINN variants is directly enforced through $\mathcal{L}_{IB}$, rather than satisfied indirectly as  in IBM (see section~\ref{sec:IBM}),  where the solid region points are required. This also suggests that MB-IBM-PINN might not be necessary when one is already aware of the solid body's position and velocity {\it a priori}. Moving forward, MB-PINN can be used efficiently for pressure recovery from velocity data even if obtained from an underlying IBM solver whenever the solid body position and velocity is known {\it a priori}. MB-PINN model is allowed here because the forcing terms and source/sink terms are zero in the fluid region and non-zero in the solid region as enforced in $\mathcal{L}_{IBvar}$ for MB-IBM-PINN. However, there might be scenarios where the solid body/ moving boundary position and velocity data might not be available {\it a priori}. This is true for applications where inferring the flow-field and the body position might be needed only from the wake data or for the prediction of an evolving moving boundary inside a closed duct/channel as in biomedical applications~\cite{weller2010free}. In such situations, one might be able to use  MB-IBM-PINN without a fluid-solid splitting of residual losses, however,  an overall physics loss relaxation would still need to be invoked.
	
	Note that the surrogate model in the present study is able to handle the immersed moving boundary using the Lagrangian markers and the fluid using an Eulerian grid, respectively. The surrogate model with residual relaxation performs very well for queries on the high resolution grids and additionally recover pressure as a hidden variable. However, the number of data points used are still large even with the coarse CI data sets with about $1.328\text{e}06$ data points in total. Therefore, given the localised features of the flow-field, the possibility of a physics based sampling of the data points is explored in the next section.

	\section{Improving data efficiency through physics-based data sampling}\label{sec:vssampling}
	
	As mentioned earlier, capturing strong/sharp gradients may become challenging in PINNs~\cite{wang2020understanding}, and hence, physics constraint relaxation was explored in section~\ref{sec:relaxation}. 
	In the baseline MB-PINN model, the secondary vortex structure was not resolved accurately while also smudging the LEV  as shown in figure~\ref{fig:baselinevorticitycontours}. This was also reflected in the locally high point-wise errors in figures~\ref{fig:baselineUVPerrcontours} (marked in rectangular boxes). These errors were localised around the regions of strong LEVs and TEVs which were representative of strong velocity gradients in the flow-field. Many authors proposed adaptive sampling strategies to mitigate this~\cite{wu2022comprehensive,nguyen2022fixed, gao2022failure, peng2022rang, tang2023pinns}. 
	It was noted that the sampling of spatial points based on a prior distribution is equivalent to point-wise weighting; ~\cite{wu2022comprehensive} dealt with PDEs in three dimensions, while~\cite{tang2023pinns} proposed a deep adaptive sampling based PINNs (DAS-PINNs) for higher dimensional PDEs.
	
	However, the above approaches entail iterative sampling of newer residual collocation points (and not data points) to mitigate the training difficulty. Recently, it was shown in~\cite{gopakumar2022loss} how the loss landscape of  PINN becomes smoother in the presence of very sparse/coarse simulation/experimental data in the bulk, as compared to no data, thereby making PINNs easier to train. In the present study, a data-driven approach was thus adopted. In fact, high number of data points were also used in the optimal MB-PINN model identified in the previous section. To improve data-efficiency and  reduce associated data storage requirement, the possibility of using the underlying features of the flow-field data to train the network for MB-PINN is explored in this section. 
	
	\begin{table}[h!]
		\centering
		\caption{Data sets generated through vorticity cutoff based sampling.}
		\begin{tabular}{cccccc}
			\hline
			Data set & $S_{\omega_z}$ & $S_{\omega_z}^{NF}$ & $N_{Bulk}$ & $N_{res}$ & $S_{Data}$ \\
			\hline
			CI & 100 & \multirow{6}{*}{100} & 1.2915e06 & \multirow{7}{*}{1.3062e06} & 4.89 \\
			CI-S50 &50 & & 7.5112e05 & & 2.85 \\
			CI-S10 & 10& & 3.1885e05 & & 1.21\\
			CI-S5 & 5 & & 2.6481e05 & & 1.00\\
			CI-S1& 1 & & 2.2158e05 & & 0.84 \\
			CI-S0& 0& & 2.1078e05 & & 0.79 \\
			& & & & &\\
			CI-S0-US10& 0 & 10 & 2.1074e04 & & 0.08\\
			\hline
		\end{tabular}
		\label{tab:vsdatasets}
	\end{table}
	\begin{figure}[t!]
		\centering
		\includegraphics[width=\linewidth]{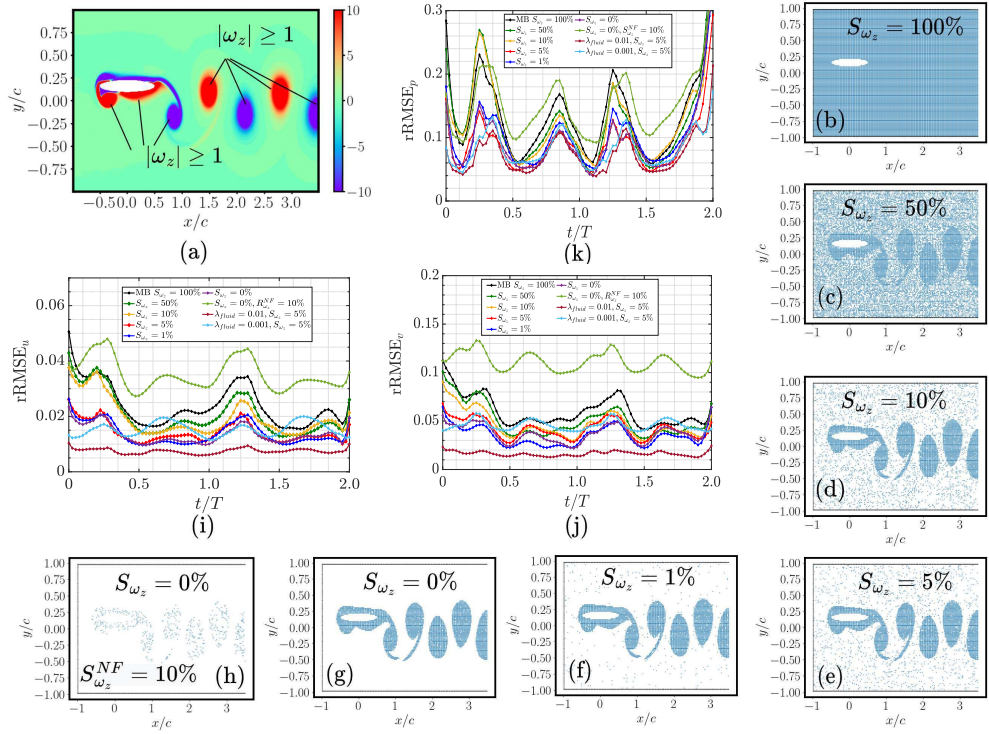}
		\caption{Vorticity cutoff based undersampling strategy: (a) the selection of strong and weak vorticity regions is shown for a  representative vorticity snapshot at $t/T = 0.0,$ which is characterised by a LEV at the bottom surface, (b)-(h) vorticity cutoff based undersampled data locations corresponding to the data sets in table~\ref{tab:vsdatasets}, (i, j) relative errors in velocity reconstruction, and (k) relative error in recovered pressure.}
		\label{fig:vssamplingstrategy}
	\end{figure}

	Vorticity magnitude levels are good indicators of strong velocity gradients and the current study proposes a data efficient vorticity cutoff based undersampling approach.  Here a vorticity cutoff $\omega_z^*$ is chosen such that the key features can be retained locally while the regions of low vorticity are undersampled. The vorticity fields are computed from the velocity data using a second order central-difference scheme. Once the vorticity fields are obtained, a reasonable absolute vorticity cutoff value of $|\omega_z^*| \; = 1.0$ is chosen. For each vorticity snapshot, only those $N_{|\omega_z|\geq|\omega_z^*|}$  fluid  points corresponding to $|\omega_z|\geq|\omega_z^*|$ are first retained, which correspond to the regions of key primary and secondary vortex structures (see figure~\ref{fig:vssamplingstrategy}(a)). Out of the remaining $N_{|\omega_z|<|\omega_z^*|}$ fluid points corresponding to $|\omega_z|<|\omega_z^*|,$ a randomly undersampled set of $N^{sample}_{|\omega_z|<|\omega_z^*|}$ points based on a percentage sampling ratio, $\displaystyle S_{\omega_z} = \frac {N^{sample}_{|\omega_z|<|\omega_z^*|}}{N_{|\omega_z|<|\omega_z^*|}} \times 100$ is constructed and combined with the earlier retained data points. Note that $|\omega_z^*|$ does not always need to be $1.0$, this may take a different value for a different flow problem and should be selected based on the flow-field vorticity levels of the specific problem of interest.
	\begin{table}[t!]
		\centering
		\caption{Accuracy details of MB-PINN models for different levels of vorticity cutoff based undersampling. If not mentioned specifically, $\lambda_{fluid}$ is taken as $\lambda_{fluid} = 0.1$.}
		\vspace{6pt}
		\resizebox{\textwidth}{!}{\begin{tabular}{cccccc}
				\hline
				Data base &$S_{\omega_z}$(in $\%$)& aRMSE &aMAE&a$R^2$&arRMSE (in $\%$) \\
				\hline
				CI &$100$& 7.64e-02	&4.09e-02	&9.91e-01	&7.27 \\
				CI-S50 &$50$&7.02e-02	&3.93e-02	&9.92e-01&	6.55 \\
				CI-S10 &$10$&6.79e-02	&3.99e-02	&9.92e-01&	6.29 \\
				\textbf{CI-S5} &$\textbf{5}$& \textbf{5.01e-02}	&\textbf{2.99e-02}	&\textbf{9.96e-01}&	\textbf{4.82} \\
				CI-S1 &$1$& 5.41e-02	&3.47e-02	&9.95e-01&	5.01\\
				CI-S0 &$0$& 5.11e-02	&3.30e-02	&9.96e-01&	4.83 \\
				CI-S0-US10 ($S^{NF}_{\omega_z}=10\%$) &$0$&9.6e-02	&4.98e-02&	9.87e-01&	9.90\\
				\textbf{CI-S5}($\lambda_{fluid} = 0.01$) &$\textbf{5}$&\textbf{ 3.59e-02}&	\textbf{2.25e-02}&	\textbf{9.97e-01}&	\textbf{3.22}\\
				CI-S5 ($\lambda_{fluid} = 0.001$)&$5$& 4.78e-02	&2.67e-02	& 9.96e-01&	4.77 \\
				\hline
		\end{tabular}}
		\label{tab:vsaccuracy}
	\end{table}
	\begin{figure}[b!]
		\centering
		\includegraphics[trim ={3cm 0 0 0}, clip, width=\linewidth]{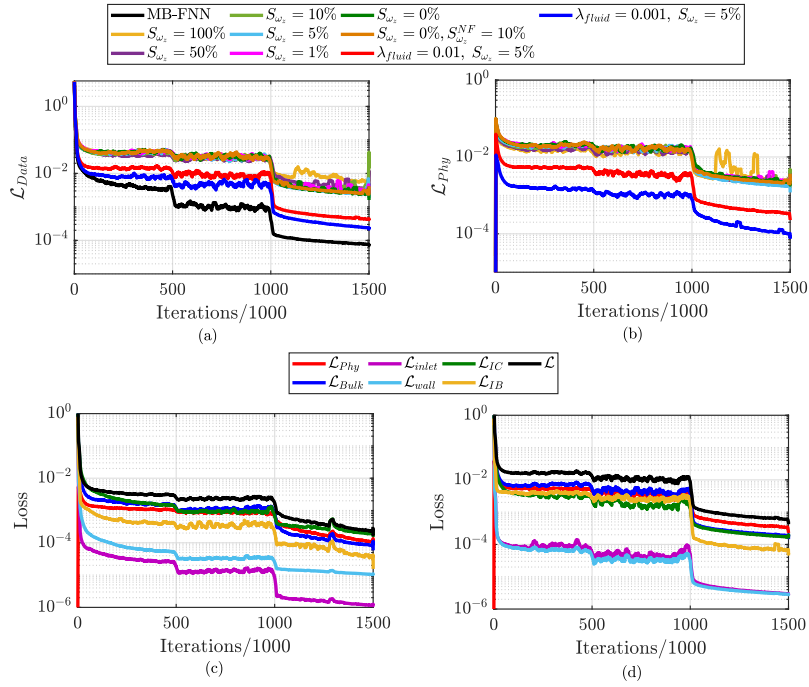}
		\caption{Loss convergence plots for select MB-PINN models considering different levels of vorticity undersampling combined with residual relaxation: (a) $\mathcal{L}_{data}$ with that of the best MB-FNN model, and (b) scaled $\mathcal{L}_{Phy}$. Unless otherwise specified, $\lambda_{fluid} = 0.1$ by default. Convergence of individual loss components for the optimal MB-PINN models with (c) $\lambda_{fluid} = 0.001,\; S_{\omega_z} = 100\%,$ and (d) $\lambda_{fluid} = 0.01,
			\;S_{\omega_z} = 5\%.$}
		\label{fig:lossvssampling}
	\end{figure}
	In the present study, $S_{\omega_z} = 100\%,50\%,10\%,5\%,1\%\text{ and } 0\%$ are considered. For the extreme case $S_{\omega_z} = 0\%$,  strong vorticity points in the near-field (NF) are further undersampled with a sampling ratio defined by, $\displaystyle S_{\omega_z}^{NF} = \frac {N^{sample}_{|\omega_z|>|\omega_z^*|}}{N_{|\omega_z|>|\omega_z^*|}} \times 100$, such that $S_{\omega_z}^{NF}=10\%.$ The spatial resolutions of the undersampled data sets are presented at a representative time stamp in figures~\ref{fig:vssamplingstrategy}(b)-(h). For each $S_{\omega_z},$ the corresponding data sets are generated  (see table~\ref{tab:vsdatasets}) with the same number of temporal snapshots as in the CI data set. The resultant models are tested on Ref-ALE for pressure recovery and Ref-IBM for velocity reconstruction. For all the cases considered here, velocity data on the boundaries, $\Gamma_{inlet}^r$, $\Gamma_{upper}^r$ and $\Gamma_{lower}^r$ (depicted by black marker points in figures~\ref{fig:vssamplingstrategy}(b)-(h)), are also used during the training step, along with the $N_{|\omega_z|\geq|\omega_z^*|}$ and $N^{sample}_{|\omega_z|<|\omega_z^*|}$ points. 
	
	To understand the effectiveness of the vorticity based undersampling, the MB-PINN models are first trained with the different data sets mentioned in table~\ref{tab:vsdatasets} for $\lambda_{fluid} = 0.1$. Here, $S_{\omega_z} = 100\%$ and $\lambda_{fluid}= 0.1$ corresponds to the baseline MB-PINN model discussed in section~\ref{sec:relaxation}. The detailed error measures are presented in table~\ref{tab:vsaccuracy}. It is observed that, all the undersampled cases except the extreme case of CI-S0-US10, perform better in terms of the arRMSE compared to the baseline MB-PINN. With $\lambda_{fluid}=0.1,$ the best results are obtained for $S_{\omega_z} = 5\%$, with an overall arRMSE of $4.82\%$ which is slightly higher than the earlier reported best case of $\lambda_{fluid} = 0.001$ (without vorticity cutoff based undersampling) in table~\ref{tab:mb-ns-ibm-pinn-accuracy} with an arRMSE of $3.217\%$. Next, keeping $S_{\omega_z} = 5\%$ fixed, as $\lambda_{fluid}$ is reduced to $0.01$, arRMSE again becomes comparable to that of $\lambda_{fluid} = 0.001$ in table~\ref{tab:mb-ns-ibm-pinn-accuracy},  that too with only $S_{Data} = 1.00\%$ for the CI-S5 data set as opposed to $4.89\%$ in the case of the CI data set. The results for $\lambda_{fluid} = 0.001$ are also reasonably good at $S_{\omega_z} = 5\%$. However, the MB-PINN model demonstrates the best performance for $\lambda_{fluid} = 0.01$ and $S_{\omega_z} = 5\%$. 
	
	Even in the snapshot-wise rRMSE plots for velocity reconstruction (figures~\ref{fig:vssamplingstrategy}(i) and~\ref{fig:vssamplingstrategy}(j)) and pressure recovery (figure~\ref{fig:vssamplingstrategy}(k)), it is seen that $\lambda_{fluid} = 0.01$ and $S_{\omega_z} = 5\%$ case performs the best overall. Interestingly, for the extreme case of $S_{\omega_z} = 0\%$ and for $S_{\omega_z}^{NF} = 10\%$, although the relative velocity reconstruction errors  worsen (figures~\ref{fig:vssamplingstrategy}(i) and~\ref{fig:vssamplingstrategy}(j)), the relative pressure errors have not  as much and are  contained below $15\%$ (figure~\ref{fig:vssamplingstrategy}(k)). This indicates that even with  $S_{Data} = 0.08\%$ with respect to Ref-IBM, MB-PINN is still able to recover pressure with reasonable accuracy. Note that both physics loss weighting and vorticity cutoff based sampling contribute towards relaxation of the physics constraints. When both are employed together, a relatively higher value of $\lambda_{fluid} = 0.01$ is found to be sufficient to match the best model performance.
	
	Hence, it is possible to leverage the underlying flow features to improve data efficiency, while maintaining the expressivity.
	This is also confirmed by the loss convergence plots in figure~\ref{fig:lossvssampling}. It can be clearly seen how the combined data-driven losses $\mathcal{L}_{Data}$ for $S_{\omega_z} = 5\%$ with $\lambda_{fluid} = 0.01, 0.001$ are closer to that of MB-FNN model in figure~\ref{fig:lossvssampling}(a) and correspondingly the scaled $\mathcal{L}_{Phy}$ are lowest in figure~\ref{fig:lossvssampling}(b). 
	The loss convergence of individual loss components for the best MB-PINN models in the relaxation and undersampling experiments are presented in figures~\ref{fig:lossvssampling}(c) and \ref{fig:lossvssampling}(d). Although the losses are slightly higher for the undersampled case, the order of convergence is still similar for both the optimal models considered. 
	
	\begin{table}[b!]
		\centering
		\caption{Accuracy of the best MB-PINN models in comparison with purely data driven MB-FNN model and the baseline MB-PINN model.}
		\vspace{6pt}
		\begin{tabular}{ccccc}
			\hline
			Model & \multicolumn{4}{c}{Accuracy} \\
			\hline
			MB-FNN & RMSE & MAE & $R^2$ & rRMSE \\
			$u$ & 5.8e-03 & 2.1e-03 & 9.999e-01 &0.51 \\
			$v$ & 6.5e-03 & 2.3e-03 & 9.998e-01 & 1.31 \\\\
			\multicolumn{5}{c}{ $\lambda_{fluid} = 0.1$ (Baseline)}\\
			MB-PINN & RMSE & MAE & $R^2$ & rRMSE \\
			$u$ & 2.7e-02	&1.1e-02	&9.994e-01	&2.41 \\
			$v$ & 3.0e-02	&1.2e-02	&9.961e-01	&6.01 \\
			$p$ & 1.7e-01 & 1.0e-01 & 9.781e-01 & 13.41 \\\\
			\multicolumn{5}{c}{ $\lambda_{fluid} = 0.001$}\\
			MB-PINN & RMSE & MAE & $R^2$ & rRMSE \\
			$u$ & 9.6e-03 &	3.8e-03	& 9.999e-01	& 0.84 \\
			$v$ & 8.3e-03 &	3.3e-03	& 9.997e-01	& 1.67 \\
			$p$ & 8.9e-02 & 	5.5e-01 & 9.943e-01	& 7.14 \\\\
			\multicolumn{5}{c}{ $\lambda_{fluid} = 0.01,$ $S_{\omega_z} = 5\%$}\\
			MB-PINN & RMSE & MAE & $R^2$ & rRMSE \\
			$u$ & 8.6e-03 &	4.9e-03 &	9.999e-01 &	0.75 \\
			$v$ & 8.2e-03 & 4.7e-03 &	9.997e-01 &	1.64 \\
			$p$ & 9.1e-02 &	5.8e-02 &	9.938e-01 &	7.28 \\
			\hline
		\end{tabular}
		\label{tab:bestmodelsaccuracy}
	\end{table}
	
	\begin{figure}[b!]
		\centering
		\includegraphics[width = 0.9\linewidth]{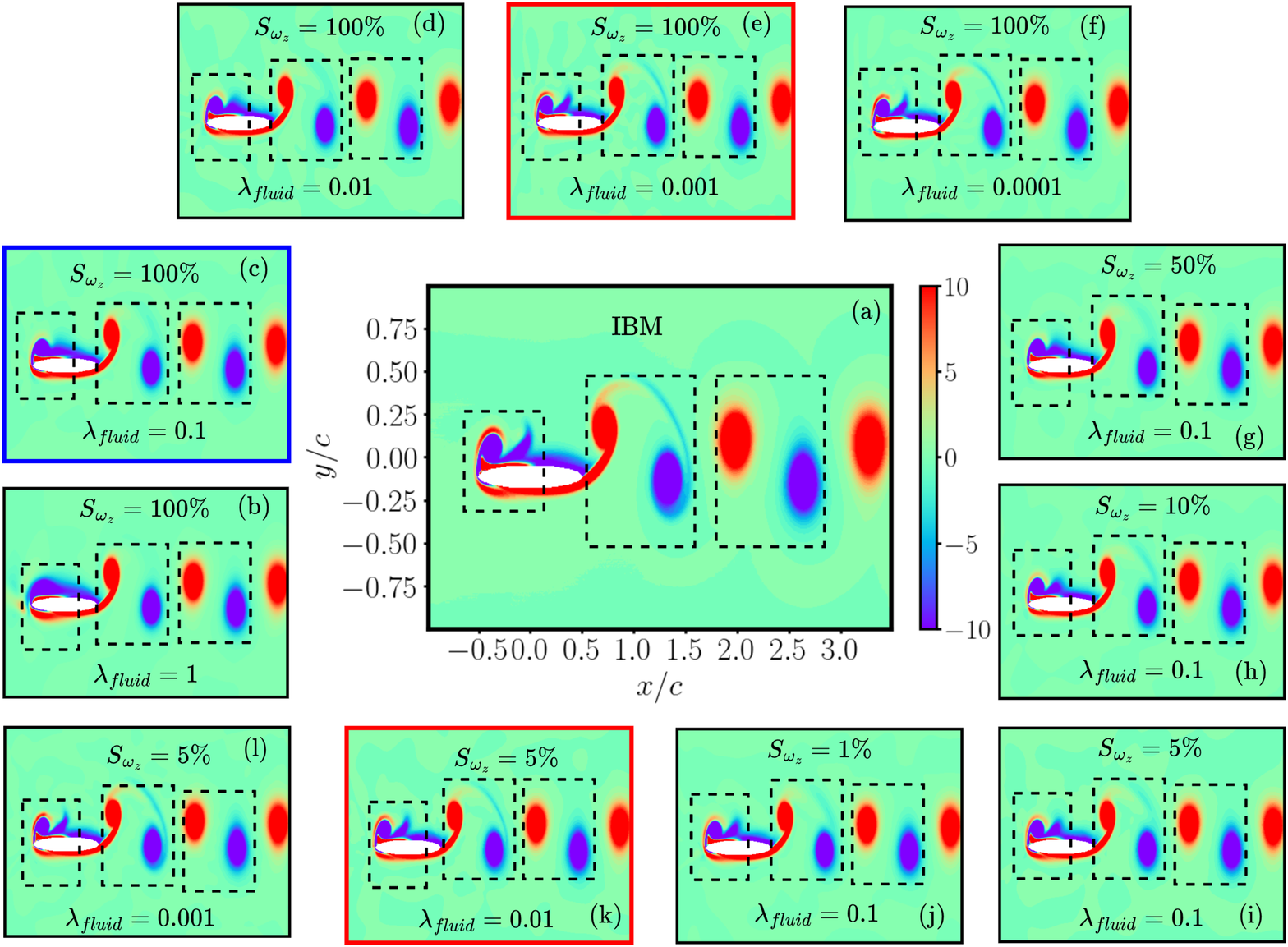}
		\caption{Comparison of vorticity contours (a) obtained from IBM data with the predictions of MB-PINN models in (b)-(l) for different $\lambda_{fluid}$ and $S_{\omega_z}$ at a typical test time stamp $t/T = 0.375.$ The dotted boxes indicate the regions of strong LEV and vortex dipoles. The base line MB-PINN model prediction in (c) is highlighted by a blue bounding box, whereas, the best MB-PINN model predictions are highlighted by a red bounding box in (f), and (l), respectively.}
		\label{fig:vorticitycontours}
	\end{figure}
	
	From the point of view of a fluid dynamicist, the accuracy of the prediction of the key vortex structures are of interest. 
	To this end, flow-field measures, such as the circulation of the LEV ($\Gamma_{LEV}$), self-induced velocity ($U_{dipole}$) of the first and the second dipoles in the wake are computed here. Errors in the predicted and derived quantities, $\Gamma_{LEV}$ and $U_{dipole}$ are good indicators of how well the primary vortex structures have been resolved by the surrogate models.
	Given a uniform grid of cell size $dx=dy$, and the vorticity $\omega_z$ at any cell center, the calculation of LEV circulation $\Gamma_{LEV}$ within the bounding box surrounding the LEV is given by
	\begin{equation}
	\Gamma_{LEV} = \sum_{i = 1}^{N_{LEV}}\omega_z^i dx dy,
	\end{equation}
	where, $N_{LEV}$ corresponds to the number of grid points in a bounding box containing the LEV. Also, the self induced velocity of a dipole A-B can be calculated as
	\begin{align}
	U_{dipole} &= \frac{\Gamma_{avg}}{2\pi\xi_{AB}},\;\;\;\mbox{where,}
	\\\Gamma_{avg} &= \frac{1}{2}(|\Gamma_A| + |\Gamma_B|), \;\;\;\mbox{and,}
	\\\xi_{AB} &= \sqrt{(x_B - x_A)^2 + (y_B - y_A)^2}.
	\end{align}
	Here, A and B indicate successive vortex cores that together form a dipole in the trailing-wake, with corresponding circulations $\Gamma_{A}$ and $\Gamma_{B}$, respectively. The dipoles are marked by the two rectangular boxes in the wake; see figure~\ref{fig:vorticitycontours}(a). $\xi_{AB}$ is the distance between the centers of vortices A and B with $(x_A,y_A) \text{ and } (x_B, y_B)$ being the respective center coordinates. For more details on the calculation of the above measures, one can refer to our earlier study~\cite{majumdar2022transition}.
	Here, the above flow-field measures are calculated at a typical test time instant, $t/T = 0.375$, which was unseen during the training. The results are presented qualitatively in figure~\ref{fig:vorticitycontours} and quantitatively in table~\ref{tab:levdipolemetrics}.
	
	\begin{table}[t!]
		\centering
		\caption{Percentage errors $\epsilon_{\#}$ in predicting LEV circulation $\Gamma_{LEV},$ self induced dipole velocity $U_{dipole}$ for the first two dipoles  in the wake at a typical test time stamp $t/T = 0.375$, for MB-PINN models. Here, $\# = \Gamma_{LEV},\, U^1, \, U^2.$ The true values of the derived quantities are $\Gamma_{LEV} = 1.450$, $U_{dipole}^1 = 0.394$, and $U_{dipole}^2 = 0.379$.}
		\begin{tabular}{cccc}
			\hline 
			Model & $\epsilon_{\Gamma_{LEV}}$ & $\epsilon_{U^1}$ & $\epsilon_{U^2}$ \\ 
			\hline 
			\multicolumn{4}{c}{Relaxation study (different $\lambda_{fluid}$)} \\ 
			\hline 
			$\lambda_{fluid}=1$ & 41.006	&0.761	&0.264 \\ 
			0.1 & 7.994	&0.309 & 0.248 \\
			0.01 & 1.033	&0.129 & 0.051 \\
			\textbf{	0.001} & \textbf{0.258	}& \textbf{0.175}&\textbf{ 0.047} \\ 
			0.0001 & 0.758	&0.132 & 0.043\\ 
			\hline 
			\multicolumn{4}{c}{Undersampling study (different $S_{\omega_z}$in $\%$))}\\ 
			\hline 
			$S_{\omega_z} = 50$& 10.425	&0.397 & 0.244 \\ 
			10& 8.139 & 0.585 & 0.188 \\ 
			5& 1.381&	0.013 &0.018 \\ 
			1& 1.968&	0.105 & 0.003 \\
			0& 1.093 & 0.016& 0.149 \\ 
			0, $S_{\omega_z}^{NF} = 10$ &3.001 & 0.399 &0.031\\
			\textbf{5}, $\lambda_{fluid} = 0.01$ & \textbf{0.073} &\textbf{0.009} &\textbf{ 0.051} \\ 
			5, 0.001 & 0.149 &	0.079 & 0.024 \\ 
			\hline 
		\end{tabular} 
		\label{tab:levdipolemetrics}
	\end{table}
	It is observed that undersampling combined with relaxation ($S_{\omega_z} = 5\%$, $\lambda_{fluid} = 0.01$) improves the LEV resolution and captures the other vortex structures remarkably well. While this is achieved, the data efficiency as compared to the optimal MB-PINN model obtained in the relaxation study, is also maintained. Notably, all the models predicted the vortex centers exactly at the locations of the respective  true data. 
	The primary and secondary vortex structures and the shear layers are resolved better compared to the baseline model  in figure~\ref{fig:vorticitycontours}(c),  either when the residuals are just further relaxed (figure~\ref{fig:vorticitycontours}(f)) or when relaxed and combined with the selective vorticity based sampling strategy (figure~\ref{fig:vorticitycontours}(l)).

	\begin{figure}[t!]
		\centering
		\includegraphics[width = 0.9\linewidth]{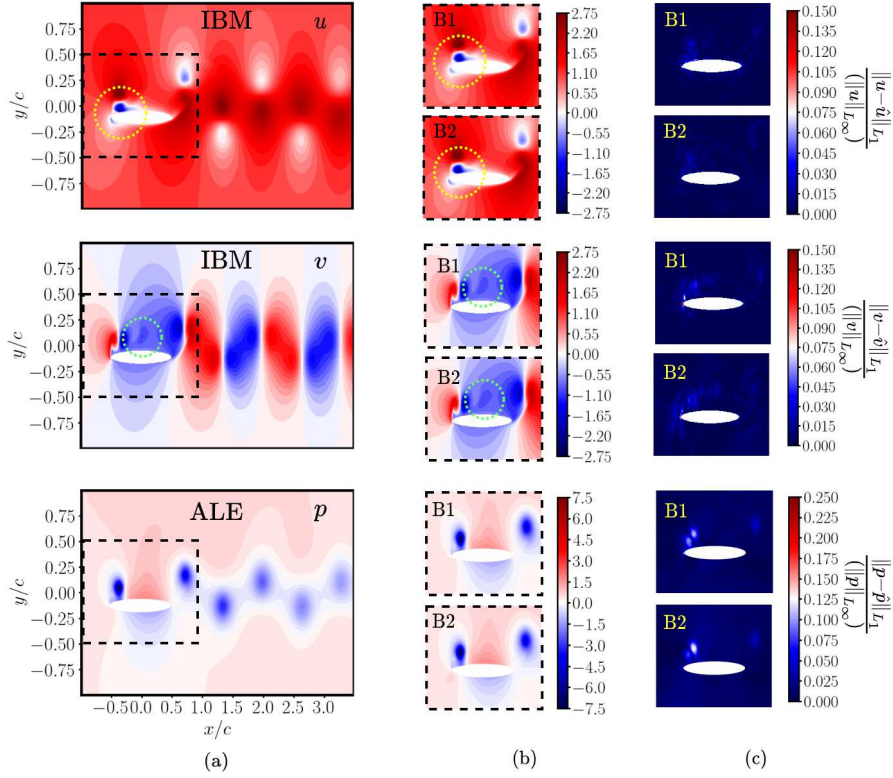}
		\caption{Comparison of (a) true IBM velocity and ALE pressure with the (b) predictions obtained from best MB-PINN models for a testing time stamp $t/T = 0.375$, and (c) corresponding maximum value normalised point wise absolute error contours. Here, the rectangular boxes in (b) and (c) representing the near-field region in (a) marked by B1 correspond to $\lambda_{fluid}$ with $S_{\omega_z} = 100\%$ while that of B2 correspond to $\lambda_{fluid} = 0.01$ with $S_{\omega_z} = 5\%.$ }
		\label{fig:bestUVPerrocntours}
	\end{figure}
	
\begin{figure}[t!]
		\centering
		\includegraphics[trim = {0 2cm 0 0}, clip, width = \linewidth]{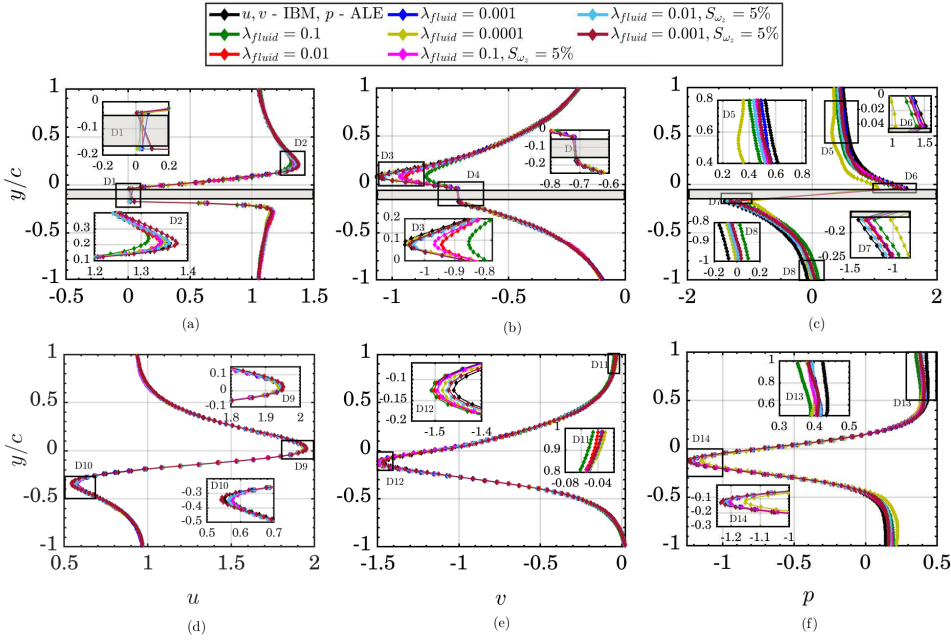}
		\caption{Velocity and pressure slices for a test time stamp $t/T = 0.375$ queried on the CI grid for different values of relaxation and vorticity undersampling. The obtained slices are compared with interpolated IBM velocity and ALE pressure data for a slice passing through, (a)-(c) the centre of the foil at $x/c = 0.0c$ where the rectangular region marked by a gray patch represents the solid body and (d)-(f) with slices at a far field location $x/c= 1.5c.$ The zoomed inshots are presented using the correspondingly labeled close up shots \textbf{D1}-\textbf{D14} }
		\label{fig:uvpslices}
	\end{figure}

	On comparing the test predictions obtained from the optimal MB-PINN models (figures~\ref{fig:bestUVPerrocntours}(a)-(c)), one can see that even the secondary structure has been captured accurately at time stamp previously unseen by the network during training. The point-wise normalised errors in pressure in the vicinity of the LEV 
	(as seen in the boxes B1 in figures~\ref{fig:bestUVPerrocntours}(b)-(c)) are slightly worse than that of $\lambda_{fluid} = 0.001$ and $S_{\omega_z}=100\%$ (as seen in the boxes B2 in figures~\ref{fig:bestUVPerrocntours}(b)-(c)). In the velocity and pressure slices (see figures~\ref{fig:uvpslices}(a)-(c)) taken at $x/c = 0.0c,$ for the test time stamp $t/T = 0.375,$ queried on the CI grid it is seen that the optimal models obtained in the relaxation and undersampling experiments are equivalent and closely match the true data. 
	Moreover it can be seen in the insets near the boundary D1 in figure~\ref{fig:uvpslices}(a), and D4 in figure~\ref{fig:uvpslices}(b) that the no-slip velocity boundary condition is closely satisfied for the optimal MB-PINN models. As seen in figure~\ref{fig:uvpslices}(c), the optimal models are very close to that of the true data in pressure recovery while for $\lambda_{fluid} = 0.0001,$ it breaks down away from the solid boundary. 
	In the slices of velocity (inshots D11-D14 of figures~\ref{fig:uvpslices}(d) and \ref{fig:uvpslices}(e)) taken at a far field location $x/c = 1.5c$ from the foil center, all the models are more or less similar in velocity reconstruction. Whereas, slight differences are seen in pressure recovery (as seen in insets D13 and D14 of figure~\ref{fig:uvpslices}(f)) across optimal and sub optimal cases though it still follows the overall trend  of the true data.  Stark differences are observed only in the near-field region  across parameters. However, it is still remarkable considering that the model was trained using an undersampling approach with about 20\% of data points compared to the CI data set and about $1\%$ to that of Ref-IBM, while still maintaining an equivalent level of performance. 
	
	Owing to the localised nature of the strong velocity gradients in the flow-field, the vorticity cutoff based data sampling leads to reasonable data efficiency (requires a significantly less number of data points for training), while maintaining the expressivity of the PINN model.
	There is almost a $5$ times decrease in the data requirement for the optimal combination of $\lambda_{fluid} = 0.01$ and $S_{\omega_z} = 5\%.$  Even at $S_{\omega_z}< 5\%,$ while  reasonable accuracy level can still be obtained. 
	It is also envisioned that, by training efficient PINN models on reduced snapshot domains of interest which have data points stored in an undersampled manner when the solver writes the data, one could potentially obtain huge savings on the memory as well. For a typical example calculation for the present problem, we observed the memory requirement per snapshot to go down by two orders of magnitude. 

	\section{Summary and conclusions}\label{sec:conclusion}
	An immersed boundary aware (body non conformal/ hybrid Lagrange-Eulerian) surrogate modeling framework has been proposed here for pressure recovery and velocity reconstruction in the unsteady flow-field past a flapping body at low Reynolds number. Physics informed neural networks are chosen as the prospective candidates. Taking inspiration from the immersed boundary method, the present framework considers a background Eulerian grid with the moving body immersed in it . This framework removes the limitations of case specific computational domain transformations to body attached frame of reference. Under the IBA framework, two formulations of PINNs, MB-PINN and MB-IBM-PINN have been presented based on the standard Navier-Stokes equations and IBM based modified formulation of the Navier-Stokes equations, respectively.  
	The present problem encompasses key characteristics of the flows past moving bodies: strong vortices with localised regions of strong gradients and time varying flow behaviour. The above mentioned characteristics pose challenges while training for PINNs in general.
	
	In the present work, the MB-PINN and MB-IBM-PINN models are specifically evaluated for velocity reconstruction and pressure field recovery as a hidden variable. To improve the model predictions under a fixed budget, a multi-part physics loss weighting strategy has been  adopted. It was discussed how, in the case where the solid body position and velocity is known {\it a priori}, 
	It has been observed that the MB-PINN approach is quite reliable where the solid body data points are discarded.  However,  MB-IBM-PINN with a suitable relative weighting of the solid region residuals, can also perform equally well  in terms of the data-driven loss components. MB-PINN outperforms MB-IBM-PINN in terms of computational time per iteration due to a relatively larger computational graph in the case of the latter. 
	Given a fixed computational training budget, it has  also been explored if the localised errors corresponding to the strong flow-field gradients in the vicinity of the solid body can be minimised and  if the error minimisation is possible with fewer data points. Towards this, a physics-based vorticity cut-off sampling strategy is adopted since the  locations with strong vorticity levels are indicators of strong gradients.
	Despite the test data set being generated from a completely different CFD solver, the best proposed MB-PINN and MB-IBM-PINN models exhibit remarkable performance in pressure recovery. It is also observed that an accurate pressure recovery is obtained irrespective of the decrease in the overall bulk data points in the training data set. 
	Hence, it is envisioned that when bulk data is available, a combination of loss components weighting with a physics based vorticity cut-off undersampling should be employed. 
	
	Although the strategies presented here require hand tuning, recent advancements in the PINNs such as adaptive training, residual collocation sampling and loss balancing methodologies could be employed in conjunction. This would limit human intervention during training while also reducing the number of runs, making the surrogate framework scalable. It would also be of interest to consider flow or kinematic parameters as additional network inputs and test the generalisability of the model on some unseen parametric instances enabling parametric differentiable surrogates.

	\bibliographystyle{elsarticle-num} 
	\bibliography{bibliography.bib}

\end{document}